\documentclass[preprint,aps,floats,nofootinbib,amssymb,superscriptaddress]{revtex4}
\usepackage[sort&compress]{natbib}
\usepackage{epsf,epsfig}
\usepackage{amsmath}
\usepackage{hyperref}
\usepackage{subfigure}
\usepackage{graphicx}
\usepackage{color}
\usepackage{mathrsfs}

\newcommand{\be}{\begin{equation}}
\newcommand{\ee}{\end{equation}}
\newcommand{\bea}{\begin{eqnarray}}
\newcommand{\eea}{\end{eqnarray}}
\newcommand{\nn}{\nonumber}
\newcommand{\ba}{\begin{array}}
\newcommand{\ea}{\end{array}}
\newcommand{\bi}{\begin{itemize}}
\newcommand{\ei}{\end{itemize}}

\begin{document}

%-----------------------------------
% Title
%-----------------------------------
\title{\vspace*{.75in}
Observing CP-violating Higgs Signals with Heavy Majorana Neutrinos}

%-----------------------------------
% Authors
%-----------------------------------

\author{Vernon Barger}\affiliation{Department of Physics, University of Wisconsin, Madison, Wisconsin 53706, USA}
\author{Wai-Yee Keung}\affiliation{Department of Physics, University of Illinois at Chicago, Chicago, Illinois 60607, USA}
\author{Brian Yencho}\affiliation{Department of Physics, University of Wisconsin, Madison, Wisconsin 53706, USA}

\thispagestyle{empty}

\begin{abstract}
\noindent We demonstrate the ability to measure the $CP$ phase of a Higgs boson state by looking at the decays of heavy, Majorana neutrinos in the process $pp \to H \to N N$ at the LHC.  This is shown for signals with both opposite-sign ($l^{\pm} l^{\mp} W^{\pm} W^{\mp}$) and same-sign ($l^{\pm} l^{\pm} W^{\mp} W^{\mp}$) leptonic final states.  These signals are investigated in the general framework of a sequential fourth generation of fermions with an additional right-handed neutrino.  Such a scenario would naturally give rise to heavy Majorana neutrinos as well as significant enhancements to Higgs production via gluon fusion due to the contributions of the new, heavy quark states running in the fermion loops.  Combined with the low background inherent to a same-sign lepton signal, this could be a useful way to investigate the $CP$ nature of a Higgs boson state at the LHC.
\end{abstract}

\maketitle

%%%%%%%%%%%%%%%%%%%%%%%%%%%%
\section{Introduction} \label{sec:intro}
%%%%%%%%%%%%%%%%%%%%%%%%%%%% 

While the Standard Model (SM) has been an extraordinarily successful theory of the strong and electroweak interactions, the precise origins of electroweak symmetry breaking are currently unknown.  In the SM, this breaking is the result of the nonzero value for the vacuum expectation value (VEV) of a single scalar field, the Higgs boson, in a process known as the Higgs Mechanism.  The SM Higgs boson has not yet been discovered, however.  Furthermore, while the existence of a single Higgs boson -- treated as a fundamental scalar particle -- may be the simplest way to break electroweak symmetry, it is not the only way nor the most attractive; the weak-scale mass of the Higgs, and its stability under quantum corrections, are unexplained in the SM if it is to be understood as part of a larger theory with mass scale $M \gg M_{W}$, such as a Grand Unified Theory (GUT).  In directly addressing these shortcomings of the SM, some theories such as weak-scale supersymmetry \cite{Baer:2006rs}, Little Higgs models \cite{ArkaniHamed:2001nc}, Technicolor \cite{Hill:2002ap}, Inert Higgs Doublet Models \cite{Barbieri:2006dq}, and Twin-Higgs Models \cite{Chacko:2005pe,Chacko:2005vw,Chacko:2005un} feature an extended Higgs(-like) sector with multiple scalar bosons.  Indeed, such extensions can be found in theories introduced to address other phenomenological issues, including the origin of dark matter \cite{McDonald:1993ex,Barger:2008jx,Goh:2009wg,Barger:2010yn} and neutrino masses \cite{Schechter:1980gr,Perez:2008ha}.

A very interesting phenomenological consequence of multi-Higgs theories is the possible introduction of new sources of $CP$ violation, which can arise from the complex mixing between the gauge and mass eigenstates of the Higgs bosons \cite{Accomando:2006ga}.  This was, in fact, one of the primary motivations for first studying the Two Higgs Doublet Model (2HDM) \cite{Lee:1973iz,Weinberg:1976hu}, one of the simplest and most generic extensions to the SM Higgs sector.  If a Higgs boson is discovered at the LHC or the Tevatron, determination of its $CP$ nature will then be vital to more fully understanding electroweak symmetry breaking; detection of a $CP$-violating phase would, for instance, be an indication of an extended sector.

A number of studies have discussed the detection of the $CP$ phase of the Higgs boson using, for instance, angular observables sensitive to $CP$ violation in the $H Z Z$ and $H W^{+} W^{-}$  couplings \cite{Chang:1993jy,Gunion:1993rs,Ilakovac:1993pt,Ma:1994kh,Gunion:1996vv,BarShalom:1997sx,Han:2000mi,Niezurawski:2004ga,Rao:2006hn,Godbole:2007cn,Christensen:2010pf} as well as in the decays of the top quark in the process $h \to t \bar{t} \to b \bar{b} W^{+} W^{-}$ \cite{Chang:1993jy,Grzadkowski:1995rx,Hasuike:1996tr,Grzadkowski:2000xs,Godbole:2002qu,Asakawa:2003dh,Khater:2003wq,Valencia:2005cx,El Kaffas:2006nt,Bhupal Dev:2007is,Hioki:2007jc,Bernreuther:2008ju,Valencia:2010dq}.  The results in the latter case can apply equally well to other heavy fermions coupling to the Higgs boson and decaying via the $W$ boson.  Indeed, extensions of the SM with a fourth generation provide such heavy, fermionic candidates, a scenario which has received renewed interest (see, for e.g., Refs.~\cite{Fargion:1999ss,Frampton:1999xi,Kribs:2007nz,Soni:2008bc,Holdom:2009rf,Hou:2010mm}).

We consider here a fourth-generation model with a weak-scale Majorana neutrino.  Through its mixing with the known fermions, it can decay to SM leptons via $N \to l^{\pm} W^{\mp}$.  The Majorana nature of these neutrinos then allows for signals with opposite-sign leptons and $W$ bosons ($H \to N N \to l^{+} l^{-} W^{+} W^{-}$) as well as same-sign ($H \to N N \to l^{\pm} l^{\pm} W^{\mp} W^{\mp}$).  The latter signal has long been understood to be the discovery mode for heavy Majorana neutrinos at colliders due to the low SM background \cite{Keung:1983uu}.  We demonstrate here that both the opposite- and same-sign signals may be used in a manner akin to the top quark signals to determine the $CP$ nature of the Higgs boson.

In Sec.~\ref{sec:masses}, we describe a heavy, fourth-generation Majorana neutrino in a simple model and give its coupling to the SM Higgs boson and electroweak gauge bosons.  We motivate the possible detection of this heavy neutrino in the $gg \to H \to N N$ mode at the LHC in Sec.~\ref{sec:production}.  We then discuss how, in the context of a more general model, the $CP$ phase of a Higgs boson can, in principle, be determined via the angular correlations of the $N$ decay products in the process $p p \to H \to N N$.

%%%%%%%%%%%%%%%%%%%%%%%%%%%%
\section{Fourth-Generation Majorana Neutrino} \label{sec:masses}
%%%%%%%%%%%%%%%%%%%%%%%%%%%%

We begin with a brief discussion of heavy Majorana neutrinos and their couplings to the Higgs boson in the context of a fourth generation\footnote{A more thorough discussion can be found in Ref.~\cite{Keung:2011zc} and references therein.  In this section we follow the conventions of Ref.~\cite{Dreiner:2008tw}.}.  We start with a simple model in which only the fourth-generation neutrino ($N$) develops mass, and therefore there is no mixing with the SM neutrinos.  Consider the following Yukawa couplings and Majorana mass terms:

\be
\mathscr{L}  \supset - Y_{n} \phi^{T} \tau L \eta_{\nu} - \frac{1}{2}M \eta_{\nu} \eta_{\nu} + \;\textrm{H.c.},
\ee

\noindent where $\phi = (\phi^{+},\phi^{0})^{T}$ is the usual SM Higgs doublet with hypercharge $Y=1/2$, $L = (\chi_{\nu}, \chi_{l})^{T}$ is the fourth-generation lepton doublet with hypercharge $Y=-1/2$, and $\eta_{\nu}$ is a SM fermion singlet with hypercharge $Y=0$.  Here, all $\chi$ and $\eta$ fields are two-component, left-handed Weyl spinors and $\tau = -i \sigma_{2}$.  After electroweak symmetry breaking, the neutral component of the Higgs doublet gets a VEV

\be
\phi \to \left( 0, v + \frac{H}{\sqrt{2}} \right)^{T},
\ee

\noindent and this part of the Lagrangian becomes

\be
\mathscr{L} \supset - Y_{n} \left(v+\frac{H}{\sqrt{2}}\right) \chi_{\nu} \eta_{\nu} - \frac{1}{2} M \eta_{\nu} \eta_{\nu} + \;\textrm{H.c.}
\ee

\noindent The mass terms can now be written as

\be
\mathscr{L} \supset - \frac{1}{2} \left( \chi_{\nu}, \eta_{\nu} \right) \bf{M_{n}} \left( \chi_{\nu}, \eta_{\nu} \right)^{T} \; + \textrm{H.c.}
\ee

\noindent with the Majorana mass matrix given by.

\be
\bf{M_{n}} = \left(
\begin{array}{cc}
0     & m_{D} \\
m_{D} & M     \\
\end{array}
\right).
\ee

\noindent Here we have defined the Dirac mass $m_{D}= Y_{n} v$.  The mass matrix $\bf{M}_{n}$ may be diagonalized by a unitary matrix $U$ to give the mass eigenstates

\be
\left(
\begin{array}{c}
N_{1} \\
N_{2} \\
\end{array}
\right)
=
\left(
\begin{array}{cc}
 i \cos\theta &  -i \sin\theta\\
   \sin\theta &     \cos\theta\\
\end{array}
\right)
\left(
\begin{array}{c}
\chi_{\nu} \\
\eta_{\nu} \\
\end{array}
\right)
\ee

\noindent with the eigenvalues

\be
M_{1,2} =  \sqrt{\left(\frac{M}{2}\right)^{2} + m_{D}^{2}} \mp \left(\frac{M}{2}\right).
\ee

\noindent When the Majorana mass term for the neutrino singlet, $M$, is much larger than $m_{D}$ we have $M_{1} \approx m_{D}^{2} / M$ and  $M_{2} \approx M$.  This is the usual see-saw mechanism invoked to explain the smallness of the masses of the known neutrinos when such a procedure is extended to the first three fermion generations (see Ref.~\cite{GonzalezGarcia:2007ib} and references therein).  For the following discussion we consider $M$ to be large enough that $N_{2}$ approximately decouples from our theory, leaving us with $\cos\theta \approx 1$ and one fourth-generation Majorana neutrino, $N_{1}$, that couples to the Higgs according to

\be
\mathscr{L} \supset - \frac{M_{1}}{\sqrt{2} v} H N_{1} N_{1} \; + \textrm{H.c.}
\ee

\noindent This can be written in the usual four-component notation with a Majorana spinor $N = (N_{1}, N_{1}^{\dagger})^{T}$ as 

\be
\label{eq:hnn}
\mathscr{L} \supset - \frac{M_{N}}{\sqrt{2} v} H \bar{N} P_{L} N \; + \textrm{H.c.},
\ee

\noindent where we have relabeled the mass as $M_{N} = M_{1}$.  In this limit, $N$ essentially acts as a gauge eigenstate when coupling to the $W$ and $Z$ bosons:

\be
\mathscr{L} \supset \frac{1}{2} \frac{g}{\cos\theta_{W}} Z_{\mu} \bar{N} \gamma^{\mu} P_{L} N + i \frac{g}{\sqrt{2}} \left(W^{+}_{\mu} \bar{N} \gamma^{\mu} P_{L} l_{4} - W^{-}_{\mu} \bar{l}_{4} \gamma^{\mu} P_{L} N \right),
\ee

\noindent where $l_{4}$ is the corresponding fourth-generation lepton whose mass is approximately $m_{D} \gg M_{N}$.  The factor of $i$ in the charged current, and relative minus sign between the two terms, is a remnant of the neutrino mixing: $\chi_{\nu} = - i \cos\theta N_{1} \approx -i N_{1}$.

%%%%%%%%%%%%%%%%%%%%%%%%%%%%
\section{Heavy Neutrino Production: $CP$-conserving Case} \label{sec:production}
%%%%%%%%%%%%%%%%%%%%%%%%%%%%

\begin{figure}[tb]
\centering
\subfigure[]{\includegraphics[clip,width=0.45\textwidth]{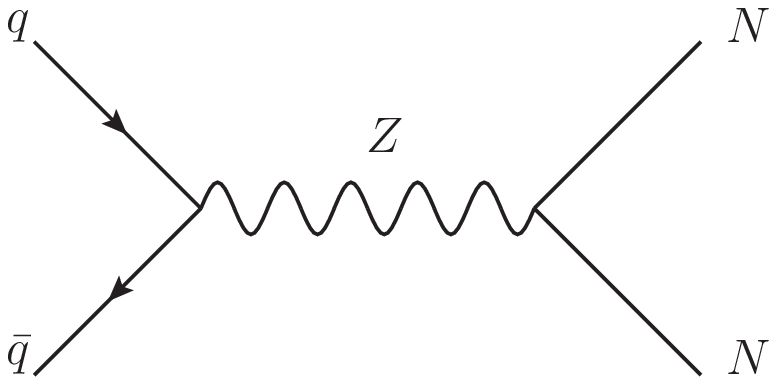}} \\
\subfigure[]{\includegraphics[clip,width=0.45\textwidth]{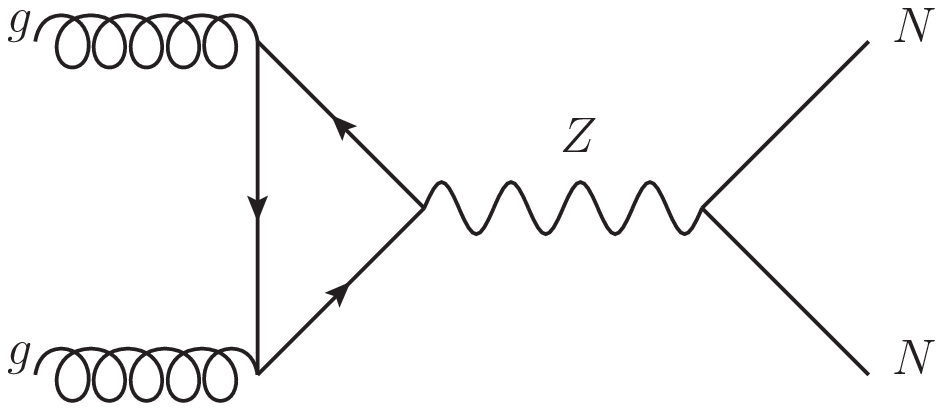}}
\subfigure[]{\includegraphics[clip,width=0.45\textwidth]{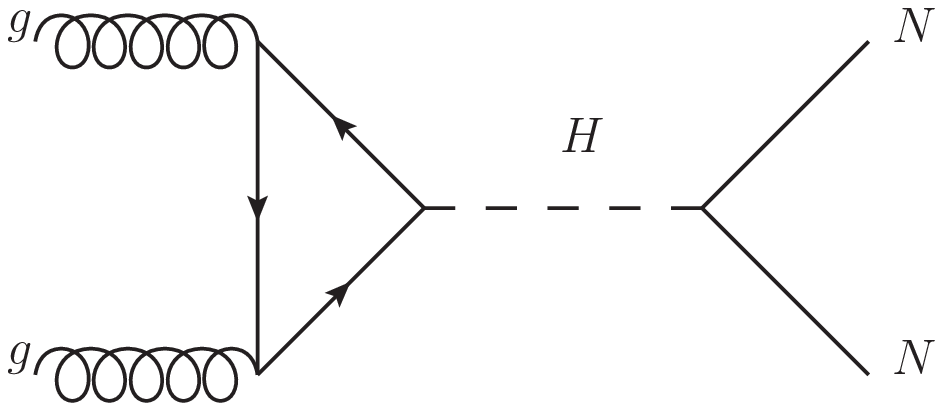}}
\caption{Dominant pair-production subprocesses of a Majorana neutrino N through (a) the Drell-Yan process and gluon fusion to (b) $Z$ and (b) Higgs bosons.}
\label{fig:diagrams}
\end{figure}

Let us now consider the pair production of such a heavy neutrino at the LHC in a model with a single, $CP$-conserving Higgs boson.  The primary production modes are the Drell-Yan process and gluon fusion to the $Z$ boson and Higgs boson via loops of heavy quarks.  The contributing diagrams are given in Figs.~\ref{fig:diagrams}~(a)-(c).  The corresponding partonic cross sections are given by

\bea
\label{eq:partonic}
\hat{\sigma}_{DY}^{q}(\hat{s}) &=& \left( \frac{\pi}{36} \right) \left( \frac{\alpha}{\sin^{2}\theta_{W} \cos^{2}\theta_{W}}\right)^{2} \left[ \left(g^{q}_{L}\right)^{2} + \left(g^{q}_{R}\right)^{2} \right] \beta_{N}^{3} \left[ \frac{\hat{s}}{\left(\hat{s}-M_{Z}^{2}\right)^{2} + \Gamma_{Z}^{2} M_{Z}^{2}}\right], \\ \nonumber
\hat{\sigma}_{Z}(\hat{s}) &=&\left(\frac{1}{1024\pi}\right) \left(\frac{\alpha \alpha_{s}}{\sin^{2}\theta_{W}} \right)^{2} \left(\frac{M_{N}}{M_{W}}\right)^{2} \left( \frac{1}{M_{W}}\right)^{2} \beta_{N} |I|^{2}, \\ \nonumber
\hat{\sigma}_{H}(\hat{s}) &=& \left(\frac{1}{2304\pi}\right) \left(\frac{\alpha \alpha_{s}}{\sin^{2}\theta_{W}} \right)^{2} \left(\frac{M_{N}}{M_{W}}\right)^{2} \left( \frac{1}{M_{W}}\right)^{2} \beta_{N}^{3} |N|^{2} \left[ \frac{\hat{s}^{2}}{\left(\hat{s}-m_{H}^{2}\right)^{2} + \Gamma_{H}^{2} m_{H}^{2}} \right], \\ \nonumber
\eea

\noindent where $v = \sqrt{2} M_{W}/g = (2\sqrt{2}G_{F})^{-1/2} \approx 174$~GeV and

\bea
\beta_{N} &=& \sqrt{1-\frac{4 M_{N}^{2}}{\hat{s}}}, \\ \nonumber
I &=& \sum_{q} 4 T_{3L}^{q} \int_{0}^{1} \rm{d}x \int_{0}^{1-x} \rm{d}y \left( \frac{x y}{x y - m_{q}^{2}/\hat{s}}\right), \\ \nonumber
N &=& \sum_{q} 3 \int_{0}^{1} \rm{d}x \int_{0}^{1-x} \rm{d}y \left( \frac{1 - 4 x y}{1 - x y \hat{s} / m_{q}^{2}}\right). \\ \nonumber
\eea

\noindent The normalized quark couplings to the $Z$ boson are given by $g^{q}_{L} = (T_{3L}^{q} - Q^{q} \sin^{2}\theta_{W})$ and $g^{q}_{R} = (- Q^{q} \sin^{2}\theta_{W})$, where $T_{3L}^{q}$ and $Q^{q}$ are the usual third component of weak isospin and electric charge, respectively, of the given quark.  We note that both $\hat{\sigma}_{Z}$ and $\hat{\sigma}_{H}$ are a factor of 2 larger than the formulas in Ref.~\cite{Willenbrock:1985tj}, in which the pair production of heavy charged leptons were studied.  This is due to the Majorana nature of the neutrinos we consider: a factor of 2 at the amplitude level from two possible Wick contractions of the neutrinos and a factor of 1/2 from identical particle phase space.

\begin{figure}[tb]
\centering
\subfigure[]{\includegraphics[clip,width=0.45\textwidth]{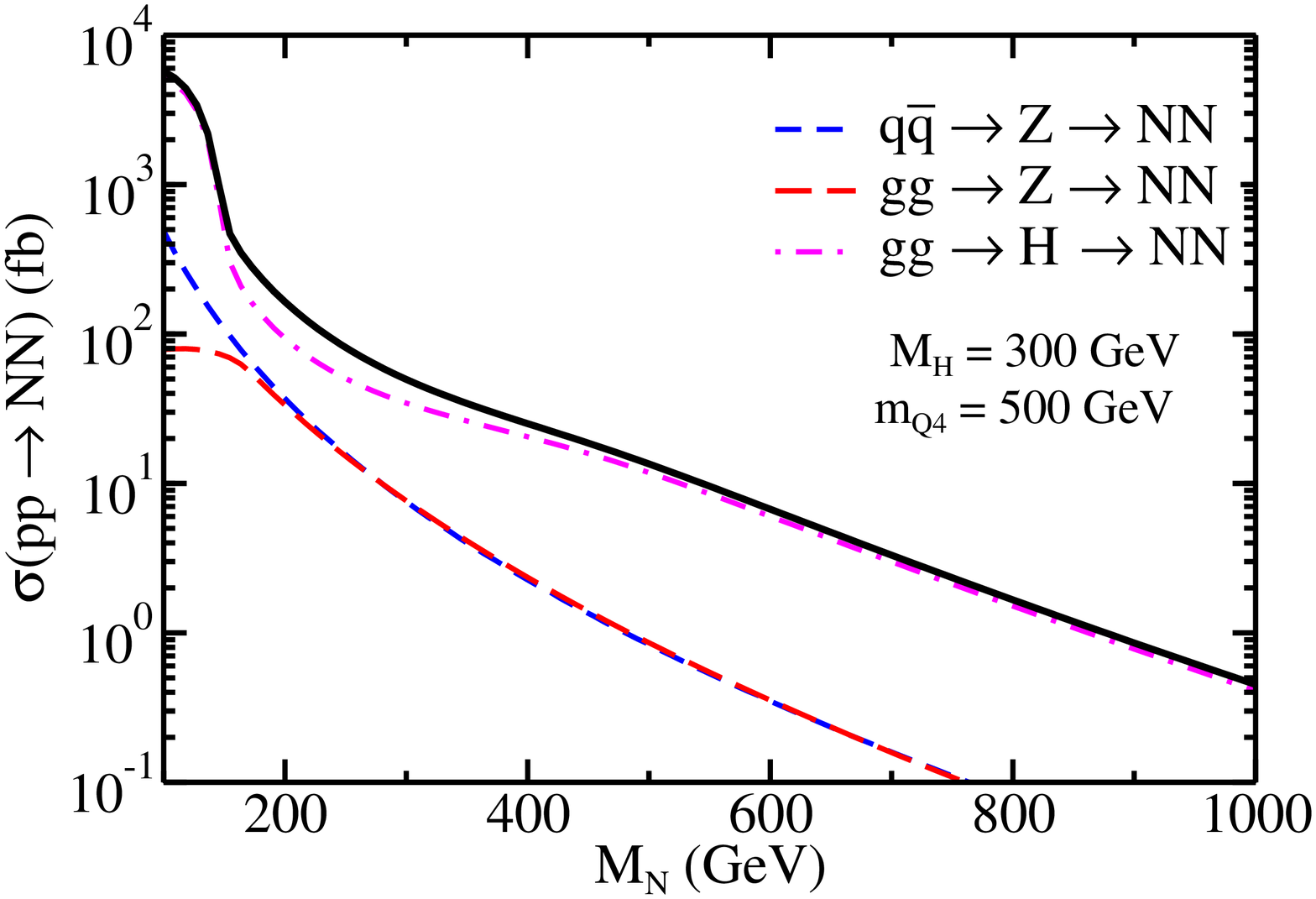}}
\subfigure[]{\includegraphics[clip,width=0.45\textwidth]{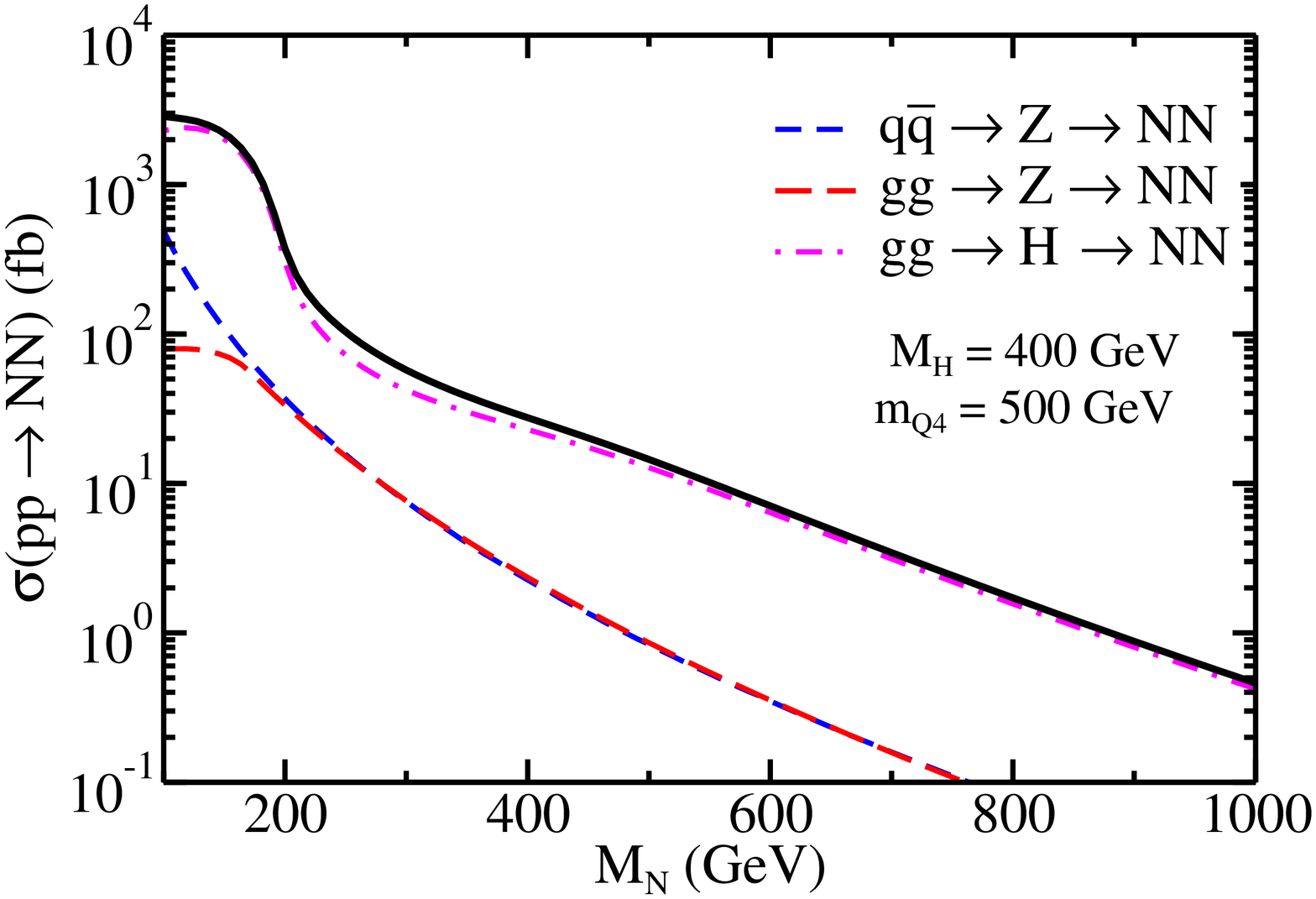}}
\subfigure[]{\includegraphics[clip,width=0.45\textwidth]{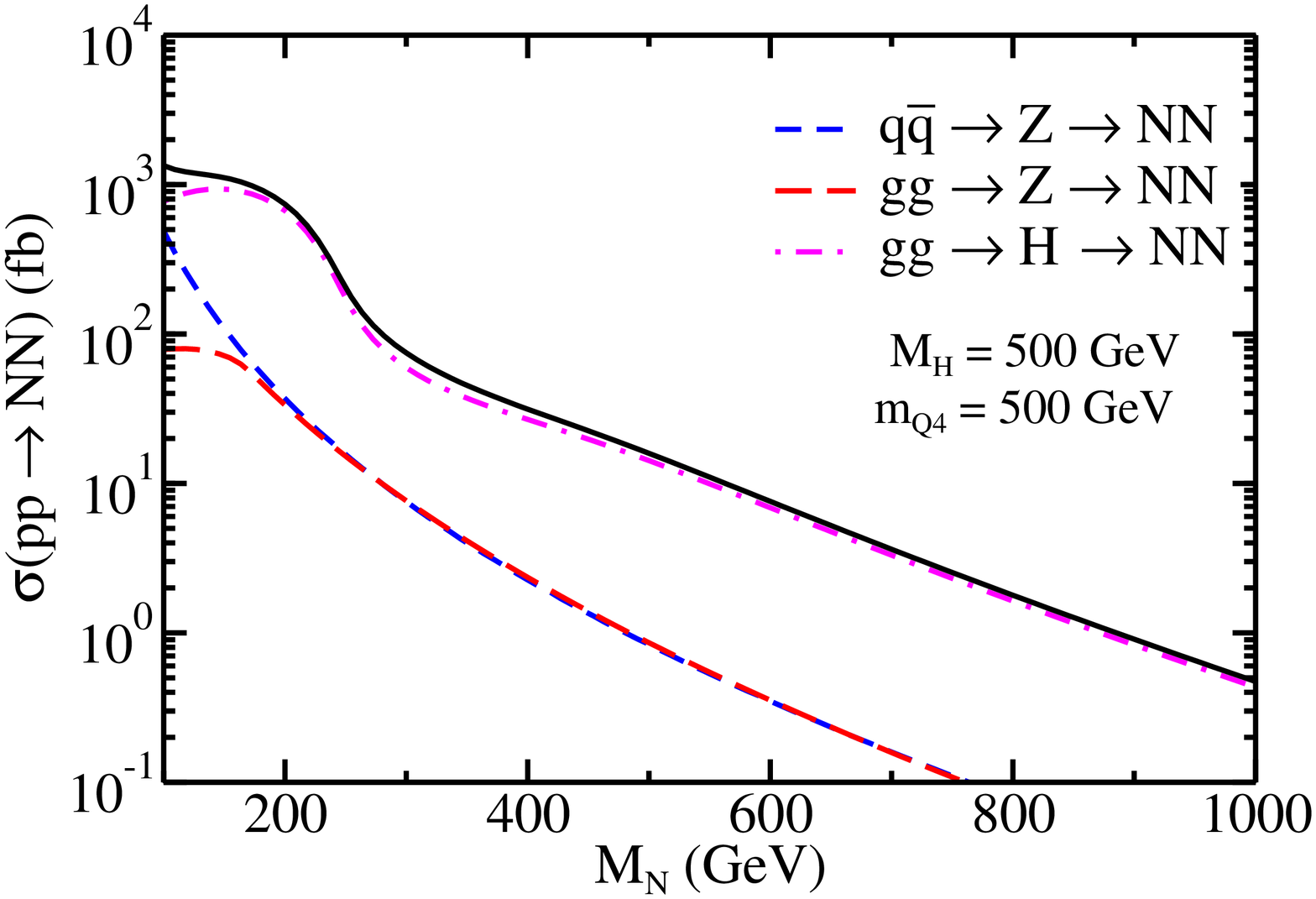}}
\subfigure[]{\includegraphics[clip,width=0.45\textwidth]{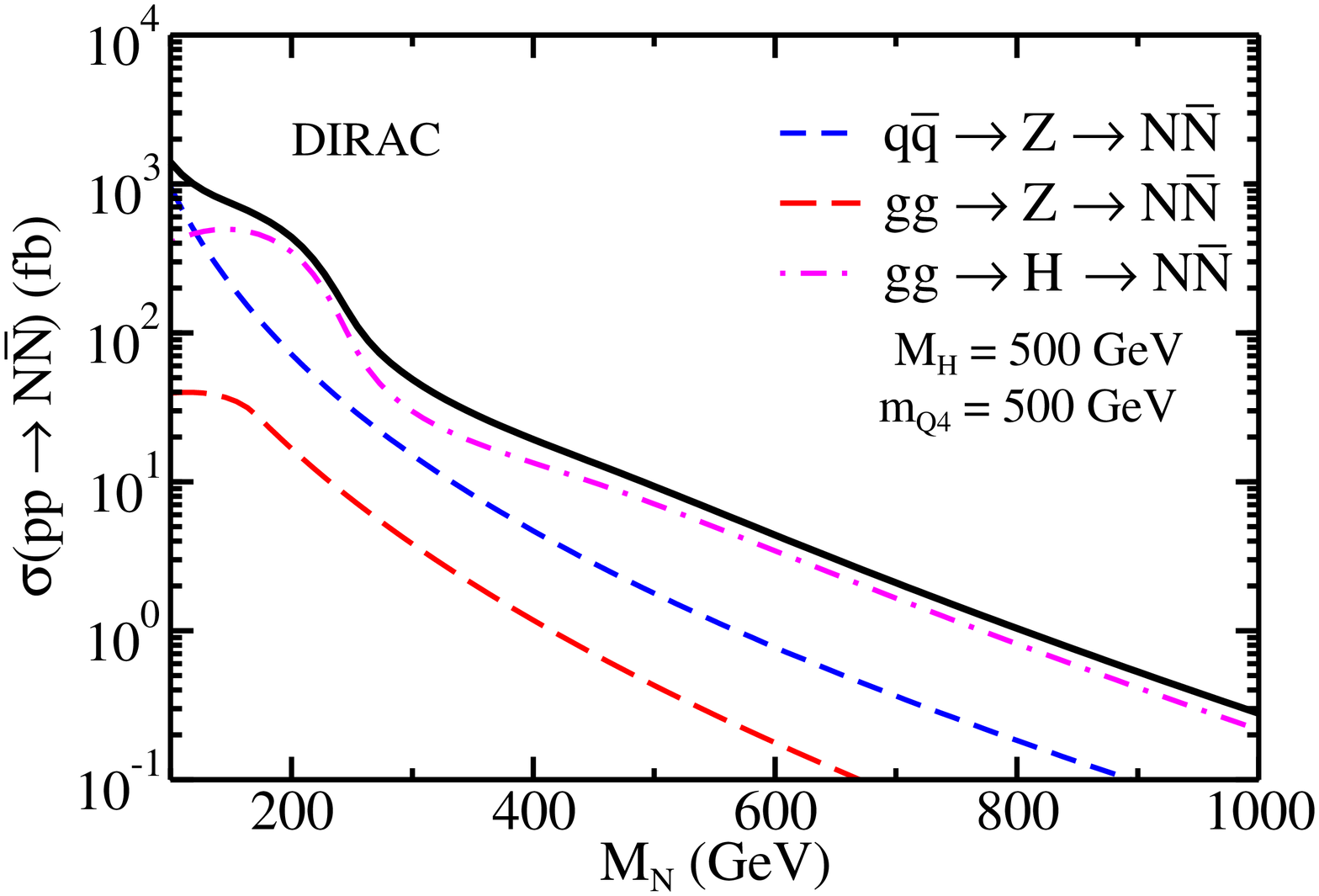}}
\caption{LHC cross sections for $pp \to NN$, the pair production of a massive neutrino of mass $M_{N}$, at $\sqrt{s}=14$~TeV with a Higgs boson mass of (a) 300 (b) 400 and (c) 500 GeV.  The results for a Dirac neutrino with a Higgs mass of 500~GeV are given in (d).  The short-dashed blue line gives the contribution from the Drell-Yan process $q\bar{q} \to Z \to N N$ , the long-dashed red line is for $g g \to Z \to N N$ and the dotted-dashed magenta line gives the contribution from $g g \to H \to N N$.  The thick black line gives the sum of the individual contributions.  In all cases the fourth-generation quark masses are set to $m_{u4}=m_{d4}=m_{Q4}=500$~GeV.}
\label{fig:production}
\end{figure}

We show the partonic cross sections in Eq.~\ref{eq:partonic} convolved with CTEQ6L1 parton distribution functions \cite{Pumplin:2002vw} for $pp$ collisions at $\sqrt{s}=14$~TeV versus the neutrino mass $M_{N}$ for three different Higgs boson masses in Figs.~\ref{fig:production}~(a)-(c).  We fix $\alpha = \alpha(M_{Z}) \approx 1/128$ and use a running value of $\alpha_{s}$ evaluated at $\mu_{R} = \mu_{F} = 2 M_{N}$, where $\mu_{F}$ is the factorization scale of our parton distribution functions and $\alpha_{s}(M_{Z})=0.118$.  For the range plotted in Figs.~\ref{fig:production}~(a)-(c), $\alpha_{s}$ varies from $\sim 0.1$ to $\sim0.08$.  Here we have chosen degenerate fourth-generation quark masses of $m_{u4}=m_{d4}=m_{Q4}=500$~GeV.  This choice, and our consideration of a fourth-generation neutrino and Higgs boson masses of several hundred GeV, are consistent with experimental constraints as well as those from electroweak precision observables in the context of a fourth generation \cite{Kribs:2007nz}.

It is clear from Figs.~\ref{fig:production}~(a)-(c) that the Higgs boson contribution is dominant in all cases.  This is largely due to the enhancement from the fourth-generation quarks\footnote{When there is a large mass splitting ($\sim$ several hundred GeV) of these quarks, the $gg \to Z \to NN$ mode can be similarly enhanced and comparable to the Higgs contribution, a case studied in Ref.~\cite{Willenbrock:1985tj} for heavy charged leptons.  However, corrections to electroweak precision parameters favor much smaller splittings, with $\Delta m_{Q4}  \approx 50$~GeV \cite{Kribs:2007nz}.  The production of heavy charged leptons from gluon fusion has since been revisited by Ref.~\cite{Liu:2010ze}, who similarly find the Higgs contribution to be dominant.  There can, however, be rather large contributions from an on-shell $Z'$ decaying to $N N$, a case studied in Ref.~\cite{delAguila:2007ua}.}, a well-known enhancement of Higgs production from gluon fusion in the context of new heavy quarks \cite{Willenbrock:1985tj,CuhadarDonszelmann:2008jp}.  In particular, in the range $M_{N} \approx 100 - 200$, when the process proceeds through a (nearly) on-shell Higgs boson, $N$ pair production can have a cross section ranging from $\sim 100$~fb to $\sim 5000$~fb.

In Fig.~\ref{fig:production}~(d) we show, for direct comparison with the Majorana case of Fig.~\ref{fig:production}~(c), the pair production of a \textit{Dirac} neutrino in a model with $m_{h} = 500$~GeV.  We note two distinct differences: the gluon fusion cross sections are a factor of 2 lower in the Dirac case, as discussed above, while the Drell-Yan process is actually enhanced.  This latter effect is the result of a higher threshold suppression found in the Majorana case due to the axial-only coupling of the heavy neutrino to the $Z$ boson.

These results suggest that if a heavy neutrino related to a fourth generation is produced in pairs at the LHC, they may be directly associated with the Higgs boson (for $M_{H} \sim$ several hundred GeV) and they may be produced at detectable rates if they are not too massive.  This link with the Higgs boson is particularly true in the Majorana case we study here, where the gluon fusion contributions are enhanced and the Drell-Yan contribution is suppressed.  We note that this assessment agrees with those of previous studies, such as Ref.~\cite{CuhadarDonszelmann:2008jp}, who performed a thorough analysis of the rates of production and decay of heavy Dirac and Majorana neutrinos at the LHC.  Similar scenarios have also been studied in Refs.~\cite{Datta:1991mf,Datta:1993nm}.

%%%%%%%%%%%%%%%%%%%%%%%%%%%%
\section{Azimuthal Distributions: $CP$-violating Case} \label{sec:phi}
%%%%%%%%%%%%%%%%%%%%%%%%%%%%

Now that we have motivated the production rates of a heavy neutrino pair in a rather basic fourth-generation model, we consider two extensions.  First, we allow small mixings with the first three generations\footnote{See, for e.g., Ref.~\cite{Hill:1989vn} in which the four generations mix and there remains at least one neutrino of weak-scale mass along with three light neutrinos.}.  This allows the heavier neutrino(s) to decay into SM particles, which we parametrize with the neutral- and charged-current interactions

\bea
\mathscr{L} & \supset & \frac{1}{2} \frac{g}{\cos\theta_{W}} \left[ U_{\nu_{l}N}^{*} \; Z_{\mu}  \bar{N} \gamma^{\mu} P_{L} \nu_{l} +  U_{\nu_{l}N} \; Z_{\mu} \bar{v}_{l} \gamma^{\mu} P_{L} N \right] \\ \nonumber
& &
+ \hspace{0.2in} \frac{g}{\sqrt{2}} \left[ V_{lN}^{*} \; W^{+}_{\mu} \bar{N} \gamma^{\mu} P_{L} l + V_{lN} \;  W^{-}_{\mu} \bar{l} \gamma^{\mu} P_{L} N \right], \nonumber
\eea

\noindent where $\nu_{l} = \nu_{e}, \nu_{\mu}, \nu_{\tau}$, and $l=e, \mu, \tau$.  $V_{lN}$ and $U_{\nu_{l}N}$ can, in principle, be different and we take each to be arbitrarily small but nonzero for at least one light lepton family.  If, as described above, we take $l_{4}$ and $h$ to be heavier than $N$, the $N \to l^{\pm} W^{\mp}$ and $N \to \nu_{l} Z$ modes are the only relevant decays and the total decay width is very narrow.  Along with very small values for $V_{lN}$ and $U_{\nu_{l}N}$, this could lead to interesting effects at the LHC such as displaced vertices or long-lived states which escape the detector and appear as large missing energy \cite{Atre:2009rg}.  However, we assume that the widths are large enough [$\Gamma_{N} >  \mathcal{O}(10^{-16})$~GeV] that the decays of the heavy neutrino are detectable inside an LHC detector.  Also, independent of the absolute size of the mixing, we note that the maximum branching fractions to opposite-sign and same-sign final states are $BR(l^{+} l^{-} W^{+} W^{-}) \approx 0.5$ and $[BR(l^{+} l^{+} W^{-} W^{-}) + BR(l^{-} l^{-} W^{+} W^{+})] \approx 0.5$, respectively, if there is only appreciable mixing to one generation and the mixing effects in the neutral current are suppressed.

The second extension to the simple model presented in the previous sections, and the one of primary interest to this study, is the possibility of an extended Higgs sector with possible $CP$ violation \cite{Accomando:2006ga}.  If, for example, we allow the Higgs boson state in Eq.~\ref{eq:hnn} to be related to the mass eigenstates by a complex mixing element such that $H = R_{1i} H_{i}$ then Eq.~\ref{eq:hnn} becomes

\be
\label{eq:hnn}
\mathscr{L} \supset - \frac{M_{N}}{\sqrt{2} v} H_{i} \bar{N} \left( R_{1i} P_{L} + R_{1i}^{*} P_{R} \right) N.
\ee

\noindent For a given $H_{i}$, the $CP$ behavior depends on the precise nature of the mixing element $R_{1i}$.  For our purposes, we will parametrize the mixing in the Higgs sector, along with any residual mixing in the neutrino sector, by a complex parameter $A$ to give the following coupling of the lightest Higgs to a heavy Majorana neutrino:

\bea
\mathscr{L}  & \supset &  - \frac{M_{N}}{\sqrt{2} v} H \bar{N} (A P_{L} + A^{*} P_{R}) N \\ \nn
                  & \supset &  - \frac{M_{N}}{\sqrt{2} v} H \bar{N} ( A_{R} - i A_{I} \gamma^{5} ) N,
\eea

\noindent where

\be
A = (A_{ R} + i A_{I}) = |A| e^{i \alpha}.
\ee

\noindent The phase of this coupling, $\alpha$, is what we are ultimately interested in determining.  Here, $\alpha=0$ corresponds to a $CP$-even Higgs state and $\alpha= \pm \pi / 2$ corresponds to a $CP$-odd Higgs state, where we restrict ourselves to $-\pi/2 < \alpha < \pi/2$.  Any nonzero value of $\alpha$ would indicate $CP$ violation.  In what follows, we seek to show that, in principle, the phase $\alpha$ may be determined by looking at the angular distributions\footnote{These angular correlations were first studied in the $CP$-conserving case in Ref.~\cite{Choudhury:1992nx}.} of the decay products of the heavy neutrinos in the process $pp \to H \to N N$.  In practice, the overall normalization of the cross section will be a model-dependent function of both the mixings in the Higgs and neutrino sectors as well as the precise nature of the Higgs' couplings to fermions.  For example, in some 2HDMs there are different Higgs bosons that give mass to the up- and down-type quarks.  Here we take a mostly model-independent approach and present results that will depend only on $M_{N}$, $M_{H}$, and $\alpha$ and are independent of the normalization of $\sigma(p p \to N N)$.  We also do not consider backgrounds, detector cuts and efficiencies, or the difficulties associated with reconstructing a boosted rest frame.  

%%%%%%%%%%%%%%%%%%%%%%%%%%%%
\subsection{Opposite-signs Case} \label{sec:os}
%%%%%%%%%%%%%%%%%%%%%%%%%%%%

Given the above definitions, the squared matrix element for $H \to NN \to l^{+} W^{-} l^{' -} W^{+}$, when summed over final-state spins, is 

\bea
\label{eq:msq_os}
\sum |\mathscr{M}|^{2} & = & (2)^{2 \theta_{M}} \left( \frac{g^{4}|V_{lN}|^{2} |V_{l'N}|^{2} M_{N}^{2}}{2} \right)
\times \left( \frac{M_{N}}{\sqrt{2} v} \right)^{2}
\times |\mathscr{P}_{n_{1}} \mathscr{P}_{n_{2}}|^{2} \times \\ \nn
& &
\bigg[ A_{I}^{2} \left( 
2 u_{1}.(n_{1}+n_{2}) u_{2}.(n_{1}+n_{2}) - (u_{1}.u_{2}) (n_{1}+n_{2})^{2}
\right) \\ \nn
& & 
+ A_{R}^{2} \left( 
2 u_{1}.(n_{1}-n_{2}) u_{2}.(n_{1}-n_{2}) - (u_{1}.u_{2}) (n_{1}-n_{2})^{2}
\right) \\ \nn
& & 
+4 \, A_{I} \, A_{R} \, \epsilon(u_{1},u_{2},n_{1},n_{2}) %\\ \nn
%& &
\bigg],
\eea

\noindent where the four-momenta are defined by

\bea
n_{1} & = &  l^{+} + W^{-}, \\ \nn
n_{2} & = &  l^{' -} + W^{+}, \\ \nn
u_{1} & = &  n_{1}-W^{-} \left(2 - \frac{(n_{1}.n_{1})}{M_{W}^{2}} \right), \\ \nn
u_{2} & = &  n_{2}-W^{+} \left(2 - \frac{(n_{2}.n_{2})}{M_{W}^{2}} \right), \\ \nn
\eea

\noindent and $l^{\pm}, W^{\pm}$ represent the corresponding four-momenta of the leptons and $W$ bosons.  We have used the shorthand notation $\epsilon(a,b,c,d) = \epsilon^{\mu \nu \rho \sigma} a_{\mu} b_{\nu} c_{\rho} d_{\sigma}$, where $\epsilon$ is the antisymmetric Levi-Civita tensor with the signature $\epsilon^{0123}=+1$.  The factors of $\mathscr{P}$ are related to the neutrino propagators and are given by

\bea
|\mathscr{P}_{n_{1}}|^{2} &=& \frac{1}{(n_{1}^{2}-M_{N}^{2})^{2} + (M_{N} \Gamma_{N})^{2}}, \\ \nn
|\mathscr{P}_{n_{2}}|^{2} &=& \frac{1}{(n_{2}^{2}-M_{N}^{2})^{2} + (M_{N} \Gamma_{N})^{2}}. \\ \nn
\eea

\noindent The factor $\theta_{M}$ is $1$ if $N$ is Majorana and is $0$ otherwise.  We note here that this formula and the results of this section apply also for $H \to t\bar{t} \to b \bar{b} W^{+} W^{-}$ if we set $\theta_{M}=0$ and replace $M_{N} \to M_{t}$, $l^{-} / l^{+} \to b / \bar{b}$, and $V_{lN} \to V_{tb}$.  In particular, Eq.~\ref{eq:msq_os} matches the results of Ref.~\cite{Chang:1993jy} when rewriting their formula in terms of on-shell $W$ bosons.

\begin{figure}[tb]
\subfigure[]{\includegraphics[clip,width=0.65\textwidth]{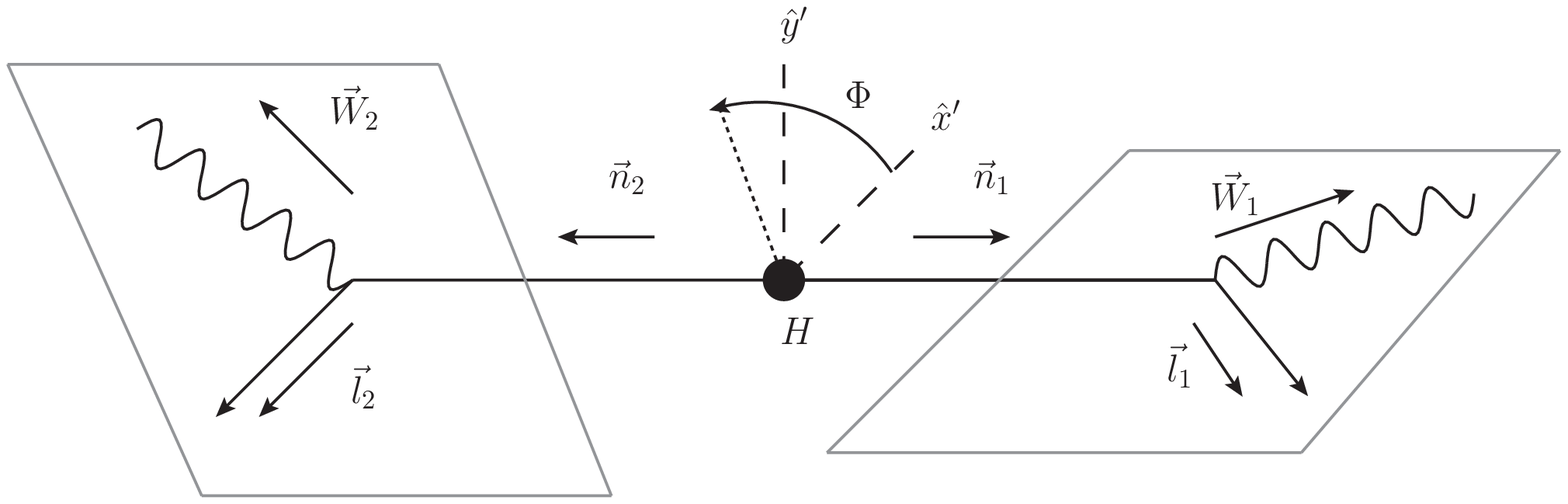}}
\subfigure[]{\includegraphics[clip,width=0.45\textwidth]{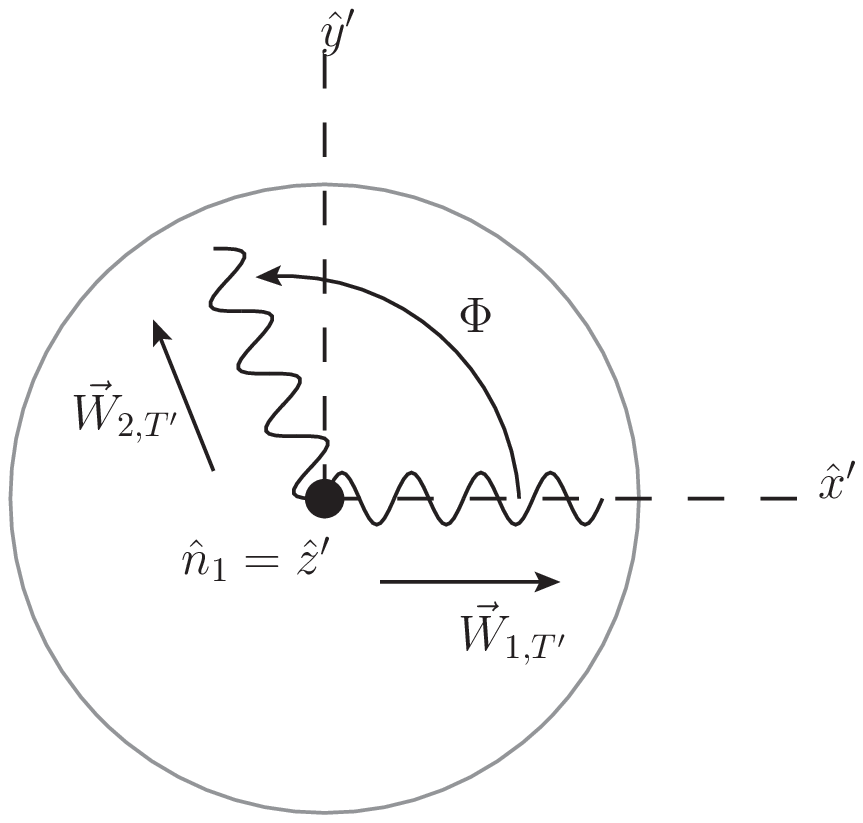}}
\caption{A schematic of the decay $H \to N N \to l^{+} l^{-} W^{+} W^{-}$ in the rest frame of the Higgs boson, with a local coordinate system defined by $\hat{z}' = \hat{n}_{1}$, the momentum direction of one of the heavy neutrinos.  Note that $\hat{z}'$ is not, generally, aligned with the beam axis (typically denoted by $\hat{z}$).  The angle $\Phi$ may be understood as the relative azimuthal angle of the decay planes of the two heavy neutrinos in the Higgs rest frame, as seen in (a).  This is also demonstrated in (b), where $\vec{W}_{1,T'}$ and $\vec{W}_{2,T'}$ denote the components of the $W$-boson momenta in a plane normal to the $\hat{z}' = \hat{n}_{1}$ axis and can be thought of as the transverse momenta of the $W$ bosons \textit{in this rest frame} (and are not to be confused with $p_{T}$).}
\label{fig:setup}
\end{figure}

We now define the $z$ axis as the $\vec{n}_{1}$ direction in the Higgs rest frame and define the sets of angles $(\theta_{1},\phi_{1})$ and $(\theta_{2},\phi_{2})$ of the $W$ bosons (or equivalently the leptons) in the rest frames of the two heavy neutrinos using a coordinate system defined with respect to this axis

\bea
\frac{\vec{W}_{1}}{|\vec{W}_{1}|} & = & (\sin\theta_{1}\cos\phi_{1},\sin\theta_{1}\sin\phi_{1},\cos\theta_{1}), \\ \nn
\frac{\vec{W}_{2}}{|\vec{W}_{2}|} & = & (\sin\theta_{2}\cos\phi_{2},\sin\theta_{2}\sin\phi_{2},\cos\theta_{2}), \\ \nn
\eea

\noindent where $W_{1}=W^{-}$ and $W_{2}=W^{+}$.  The \textit{relative azimuthal angle} of the $W$ bosons/leptons with respect to the $\vec{n}_{1}$ axis is $\Phi = (\phi_{2}-\phi_{1})$.  This is the primary observable of interest and is similar to that proposed in studies of $CP$ correlations in $h \to t\bar{t} \to b \bar{b} W^{+} W^{-}$ (see, for e.g., Ref.~\cite{Grzadkowski:1995rx}) and generic studies of $CP$-violating observables \cite{Han:2009ra}.  This geometry is illustrated in Figs.~\ref{fig:setup}~(a)-(b).

After making the narrow width approximation for $N$ and integrating over all angles except $\Phi$, we obtain

\bea
\frac{d \Gamma}{d \Phi} & \propto &  [ \, 16 (M_{N}^{2} + 2 M_{W}^{2})^{2} \, (\beta_{N}^{2} A_{R}^{2} 
+ A_{I}^{2}) \\ \nn
& & 
\, - \pi^{2} (M_{N}^{2} - 2 M_{W}^{2})^{2} ( ( \beta_{N}^{2} A_{R}^{2} - A_{I}^{2}) \cos\Phi + 2 \beta_{N} A_{R} A_{I} \sin\Phi) ]
\eea

\noindent for the process $H \to NN \to l^{+} W^{-} l^{'-} W^{+}$.  Here $\beta_{N} = \sqrt{1 - (4 M_{N}^{2}/ M_{H}^{2})}$ is the velocity of the heavy neutrino in the Higgs rest frame, taken here to be on shell.  As discussed above, the decay width of the neutrino should be extremely small, and therefore the use of the narrow width approximation here is very well justified\footnote{There are certain special cases, such as when $M_{N} \approx M_{W}$, for which the narrow width approximation for $N$ may actually give large deviations from the true result even when $(\Gamma_{N}/M_{N})$ is small.  We defer the reader to Ref.~\cite{nwa} for further details.}.

The preceding equation can be written in a normalized form as

\bea
\frac{1}{\Gamma} \frac{d \Gamma}{d \Phi} & = &  \frac{1}{2 \pi} \left[ 1 - \frac{\pi^{2}}{16} \frac{(M_{N}^{2} - 2 M_{W}^{2})^{2}}{(M_{N}^{2} + 2 M_{W}^{2})^{2}} \left( \frac{(\beta_{N}^{2} A_{R}^{2} - A_{I}^{2})}{(\beta_{N}^{2} A_{R}^{2} + A_{I}^{2})} \cos\Phi + 2 \frac{\beta_{N} A_{R} A_{I}}{(\beta_{N}^{2} A_{R}^{2} + A_{I}^{2})} \sin\Phi \right) \right ].
\eea

\noindent Finally, we define a new angle $\chi$ according to

\be
(\beta_{N} A_{R} \pm i A_{I}) = (\beta_{N}^{2} A_{R}^{2} + A_{I}^{2})^{\frac{1}{2}} e^{\pm i \chi}
\ee

\noindent that is related to $\alpha$ (the $CP$ phase of the Higgs boson) by 
\be
\chi = \tan^{-1}\left(\frac{\tan{\alpha}}{\beta_{N}}\right).
\ee

\noindent The $\Phi$ distribution can now be given as

\bea
\label{eq:phi_os}
\frac{1}{\Gamma} \frac{d \Gamma}{d \Phi} & = &  \frac{1}{2 \pi} \left[ 1 - \left(\frac{\pi}{4}\right)^{2} \frac{\left(1 - 2 \frac{M_{W}^{2}}{M_{N}^{2}}\right)^{2}}{\left(1 + 2 \frac{M_{W}^{2}}{M_{N}^{2}}\right)^{2}} \cos(\Phi - 2 \chi) \right ].
\eea

\begin{figure}[tb]
\subfigure[]{\includegraphics[clip,width=0.45\textwidth]{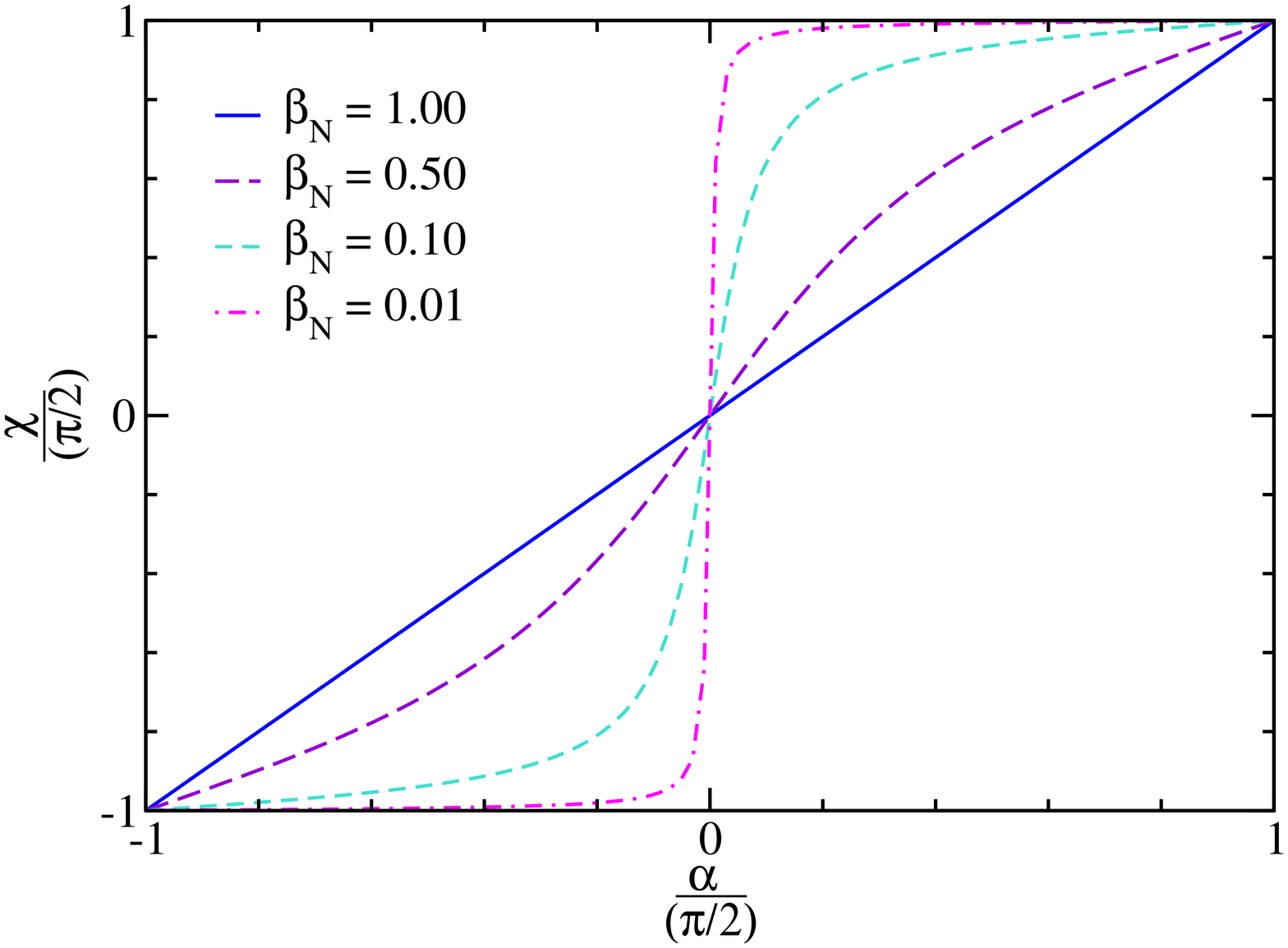}}
\subfigure[]{\includegraphics[clip,width=0.45\textwidth]{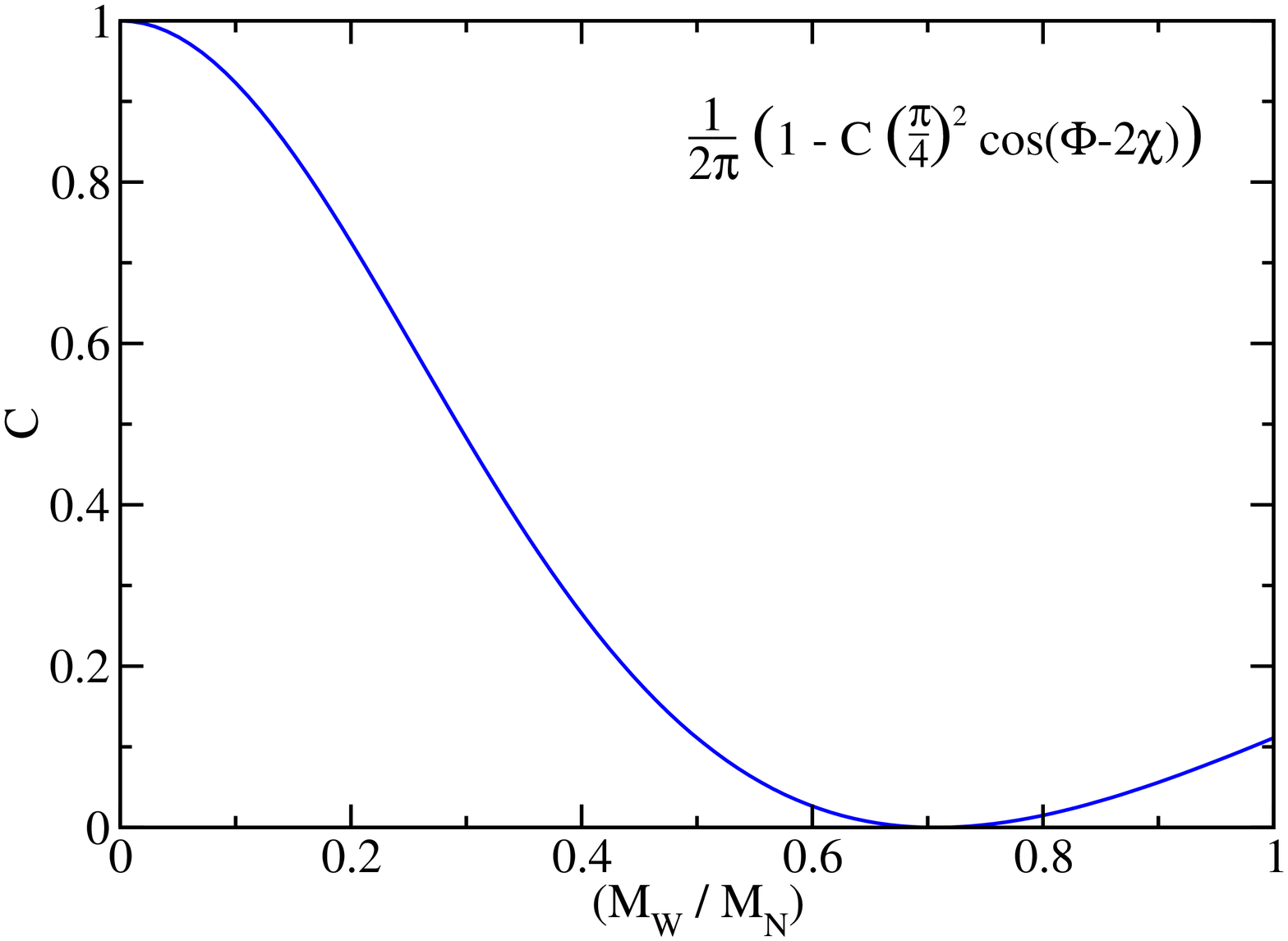}}
\caption{(a) The phase angle $\chi$ as a function of $\alpha$ for several different values of $\beta_{N} = \sqrt{1-4 m_{N}^{2} / m_{H}^{2}}$. (b) The normalized, relative amplitude $C$ of the $\Phi$ dependence of the differential partial amplitude as a function of the ratio $M_{W}/M_{N}$.  Here, $C=\left(1 - 2 \frac{M_{W}^{2}}{M_{N}^{2}}\right)^{2} / \left(1 + 2 \frac{M_{W}^{2}}{M_{N}^{2}}\right)^{2}$.} \qquad \\
\label{fig:dist}
\end{figure}

\noindent The overall effect of the $CP$ phase $\alpha$ in the coupling of the Higgs boson to $N N$ is to introduce a phase shift in the $\cos(\Phi)$ dependence of the differential decay width.  In the limit $M_{H}^{2} \gg 4 M_{N}^{2}$, we find $\chi \to \alpha$ and the total phase shift is approximately $2 \alpha$, providing a sensitive probe of the $CP$ nature of the Higgs boson.  In the other limit, $M_{H}^{2} \approx 4 M_{N}^{2}$ and the $N N$ are produced just above threshold.  This leads to $\chi \approx \pm \pi/2$ for all but the smallest values of $\alpha$.  This behavior is shown in Fig.~\ref{fig:dist}~(a).  Measuring $\alpha$ in this regime would therefore require knowing $\beta_{N}$ to very high accuracy.

In general, the measurement of this phase will depend on the ability to detect the actual $\Phi$ dependence of the distribution and, therefore, on the overall amplitude of the oscillation about the mean value.  This is a function solely of the ratio $M_{W}/M_{N}$ and is plotted in Fig.~\ref{fig:dist}~(b).  In the limit that $M_{N} \gg M_{W}$, the amplitude (relative to the mean value) reaches a maximum value of approximately $(\pi/4)^{2} \approx 0.6$.  On the other hand, the amplitude vanishes for $M_{N}=\sqrt{2}M_{W}$.  Therefore, the $CP$ effects could be difficult to observe if $N$ is not much heavier than the $W$ boson.

For completeness, we also give the result for $H \to t\bar{t} \to b \bar{b} W^{+} W^{-}$:

\bea
\label{eq:phi_top}
\frac{1}{\Gamma} \frac{d \Gamma}{d \Phi} & = &  \frac{1}{2 \pi} \left[ 1 - \left(\frac{\pi}{4}\right)^{2} \frac{\left(1 - 2 \frac{M_{W}^{2}}{M_{t}^{2}}\right)^{2}}{\left(1 + 2 \frac{M_{W}^{2}}{M_{t}^{2}}\right)^{2}} \cos(\Phi - 2 \chi_{t}) \right ],
\eea

\noindent where

\be
\chi_{t} = \tan^{-1}\left( \frac{\tan\alpha}{\beta_{t}}\right) = \tan^{-1}\left( \frac{\tan\alpha}{\sqrt{1-4 M_{t}^{2}/m_{H}^{2}}}\right).
\ee

\noindent This is the same as Eq.~\ref{eq:phi_os} with the replacement $M_{N} \to M_{t}$.  Indeed, for any heavy fermion of mass $M_{F}$ coupling to the Higgs boson in a manner similar to Eq.~\ref{eq:hnn} and decaying via $F \to f' W^{\pm}$, this formula will apply with the simple replacement $M_{N} \to M_{F}$ independent of both $BR(H \to FF)$ and $BR(F \to f' W^{\pm})$.

%%%%%%%%%%%%%%%%%%%%%%%%%%%%
\subsection{Same-sign Case} \label{sec:ss}
%%%%%%%%%%%%%%%%%%%%%%%%%%%%

While the opposite-sign signal has broad applicability to heavy fermions, we will now show a similar signal that is unique to Majorana fermions.  The squared matrix element for $H \to NN \to l^{\pm} l^{'\pm} W^{\mp} W^{\mp}$, when summed over final-state spins, is given by

\bea
\sum |\mathscr{M}|^{2} & = & - (2)^{2} \left( \frac{g^{4}|V_{lN}|^{2} |V_{l'N}|^{2} M_{N}^{2}}{2} \right)
\times \left( \frac{M_{N}}{\sqrt{2} v} \right)^{2}
\times |\mathscr{P}_{n_{1}} \mathscr{P}_{n_{2}}|^{2} \times \\ \nn
& &
\bigg[ A_{I}^{2} \bigg\{ 
2 u_{1}.(n_{1}+n_{2}) u_{2}.(n_{1}+n_{2}) - (u_{1}.u_{2}) (n_{1}+n_{2})^{2}
\\ \nn & &  \qquad
-4(u_{1}.n_{1})(u_{2}.n_{2})\left(1+\frac{(n_{1}.n_{2})}{M_{N}^{2}}\right)
\\ \nn & &  \qquad
-2(u_{1}.n_{2})(u_{2}.n_{2})\left(1-\frac{(n_{1}.n_{1})}{M_{N}^{2}}\right)
-2(u_{1}.n_{1})(u_{2}.n_{1})\left(1-\frac{(n_{2}.n_{2})}{M_{N}^{2}}\right)
\\ \nn & &  \qquad
-M_{N}^{2}(u_{1}.u_{2})\left(1-\frac{(n_{1}.n_{1})}{M_{N}^{2}}\right)\left(1-\frac{(n_{2}.n_{2})}{M_{N}^{2}}\right)
\bigg\} \\ \nn
& & 
+ A_{R}^{2} \bigg\{
2 u_{1}.(n_{1}-n_{2}) u_{2}.(n_{1}-n_{2}) - (u_{1}.u_{2}) (n_{1}-n_{2})^{2}
\\ \nn & &  \qquad
+4(u_{1}.n_{1})(u_{2}.n_{2})\left(1-\frac{(n_{1}.n_{2})}{M_{N}^{2}}\right)
\\ \nn & &  \qquad
-2(u_{1}.n_{2})(u_{2}.n_{2})\left(1-\frac{(n_{1}.n_{1})}{M_{N}^{2}}\right)
-2(u_{1}.n_{1})(u_{2}.n_{1})\left(1-\frac{(n_{2}.n_{2})}{M_{N}^{2}}\right)
\\ \nn & &  \qquad
-M_{N}^{2}(u_{1}.u_{2})\left(1-\frac{(n_{1}.n_{1})}{M_{N}^{2}}\right)\left(1-\frac{(n_{2}.n_{2})}{M_{N}^{2}}\right)
\bigg\} \\ \nn
& & 
+4 \, A_{I} \, A_{R} \, \epsilon(u_{1},u_{2},n_{1},n_{2}) %\\ \nn
%& &
\bigg]
\\ \nn & &
+ (W_{1} \leftrightarrow W_{2}).
\eea

\noindent Here we have ignored any interference with the crossed diagram, whose amplitude-squared contribution is represented here by the $(W_{1} \leftrightarrow W_{2})$ term.  This should be an excellent approximation, as the decay width of the Majorana neutrino should be extremely narrow.  We can simplify the matrix element squared further if we only keep the terms that remain in the narrow width approximation:

\bea
\label{eq:msq_ss}
\sum |\mathscr{M}|^{2} & = & - (2)^{2} \left( \frac{g^{4}|V_{lN}|^{2} |V_{l'N}|^{2} M_{N}^{2}}{2} \right)
\times |\mathscr{P}_{n_{1}} \mathscr{P}_{n_{2}}|^{2} \times \\ \nn
& &
\bigg[ A_{I}^{2} \bigg\{ 
2 u_{1}.(n_{1}+n_{2}) u_{2}.(n_{1}+n_{2}) - (u_{1}.u_{2}) (n_{1}+n_{2})^{2}
\\ \nn & & \qquad
-4(u_{1}.n_{1})(u_{2}.n_{2})\left(1+\frac{(n_{1}.n_{2})}{M_{N}^{2}}\right)
\bigg\} \\ \nn
& & 
+ A_{R}^{2} \bigg\{
2 u_{1}.(n_{1}-n_{2}) u_{2}.(n_{1}-n_{2}) - (u_{1}.u_{2}) (n_{1}-n_{2})^{2}
\\ \nn & & \qquad
+4(u_{1}.n_{1})(u_{2}.n_{2})\left(1-\frac{(n_{1}.n_{2})}{M_{N}^{2}}\right)
\bigg\} \\ \nn
& & 
+4 \, A_{I} \, A_{R} \, \epsilon(u_{1},u_{2},n_{1},n_{2}) %\\ \nn
%& &
\bigg]
\\ \nn & &
+ (W_{1} \leftrightarrow W_{2}).
\eea

\noindent There are some notable differences between Eq.~\ref{eq:msq_ss} and the opposite-sign case in Eq.~\ref{eq:msq_os}.  Apart from the $(W_{1} \leftrightarrow W_{2})$ term, there are terms proportional to $(u_{1}.n_{1})(u_{2}.n_{2})$ that are absent from Eq.~\ref{eq:msq_os}.  In addition, the terms that Eqs.~\ref{eq:msq_ss} and Eq.~\ref{eq:msq_os} have in common differ by an overall minus sign.  These differences ultimately lead to different behavior in the $\Phi$ distribution.

If we now label the leptons and $W$ bosons, the four-momenta are given by

\bea
n_{1} & = &  l_{1} + W_{1}, \\ \nn
n_{2} & = &  l_{2} + W_{2}, \\ \nn
u_{1} & = &  n_{1}-W_{1} \left(2 - \frac{(n_{1}.n_{1})}{M_{W}^{2}} \right), \\ \nn
u_{2} & = &  n_{2}-W_{2} \left(2 - \frac{(n_{2}.n_{2})}{M_{W}^{2}} \right). \\ \nn
\eea

\noindent The crossed term simply swaps the $W$ boson labels in these definitions (e.g. $n_{1} \to l_{1} + W_{2}$).  In the discussion that follows, we can effectively ignore this term; it gives no contribution to the overall decay width because of the compensating symmetry factor for the identical bosons, and any contribution to the angular distributions can be recast in terms of the first term by a relabeling of the particle states.

If we now define $\Phi$ and $\chi$ as before, we find that the azimuthal distribution for the $W$ bosons/leptons is given by

\bea
\frac{1}{\Gamma} \frac{d \Gamma}{d \Phi} & = &  \frac{1}{2 \pi} \left[ 1 + \left(\frac{\pi}{4}\right)^{2} \frac{\left(1 - 2 \frac{M_{W}^{2}}{M_{N}^{2}}\right)^{2}}{\left(1 + 2 \frac{M_{W}^{2}}{M_{N}^{2}}\right)^{2}} \cos(\Phi - 2 \chi) \right ].
\eea

\noindent This differs from the result of Eq.~\ref{eq:phi_os} for the opposite-sign case only in the relative sign between the constant and $\cos(\Phi-2\chi)$ terms.  It is therefore the same distribution shifted by an additional phase of $\pi$.  That the same-sign signal also shows a $\cos(\Phi-2\chi)$-dependence in the differential partial width is of particular interest: this signal will have considerably less SM background than the opposite-sign case and will therefore be much easier to detect and reconstruct.  In addition, when adding the $H \to NN \to l^{+}l^{+}W^{-}W^{-}$ and $H \to N N \to l^{-}l^{-}W^{+}W^{+}$ contributions together, the overall rate should be the same as in the opposite-sign case.  The same-sign signal would therefore be much better suited for making a determination of the factor $2 \chi$ and, hence, $\alpha$ and the $CP$ nature of the Higgs boson.

%%%%%%%%%%%%%%%%%%%%%%%%%%%%
\section{Comparison to MadGraph} \label{sec:madgraph}
%%%%%%%%%%%%%%%%%%%%%%%%%%%%

\begin{figure}[tb]
\subfigure[]{\includegraphics[clip,width=0.45\textwidth]{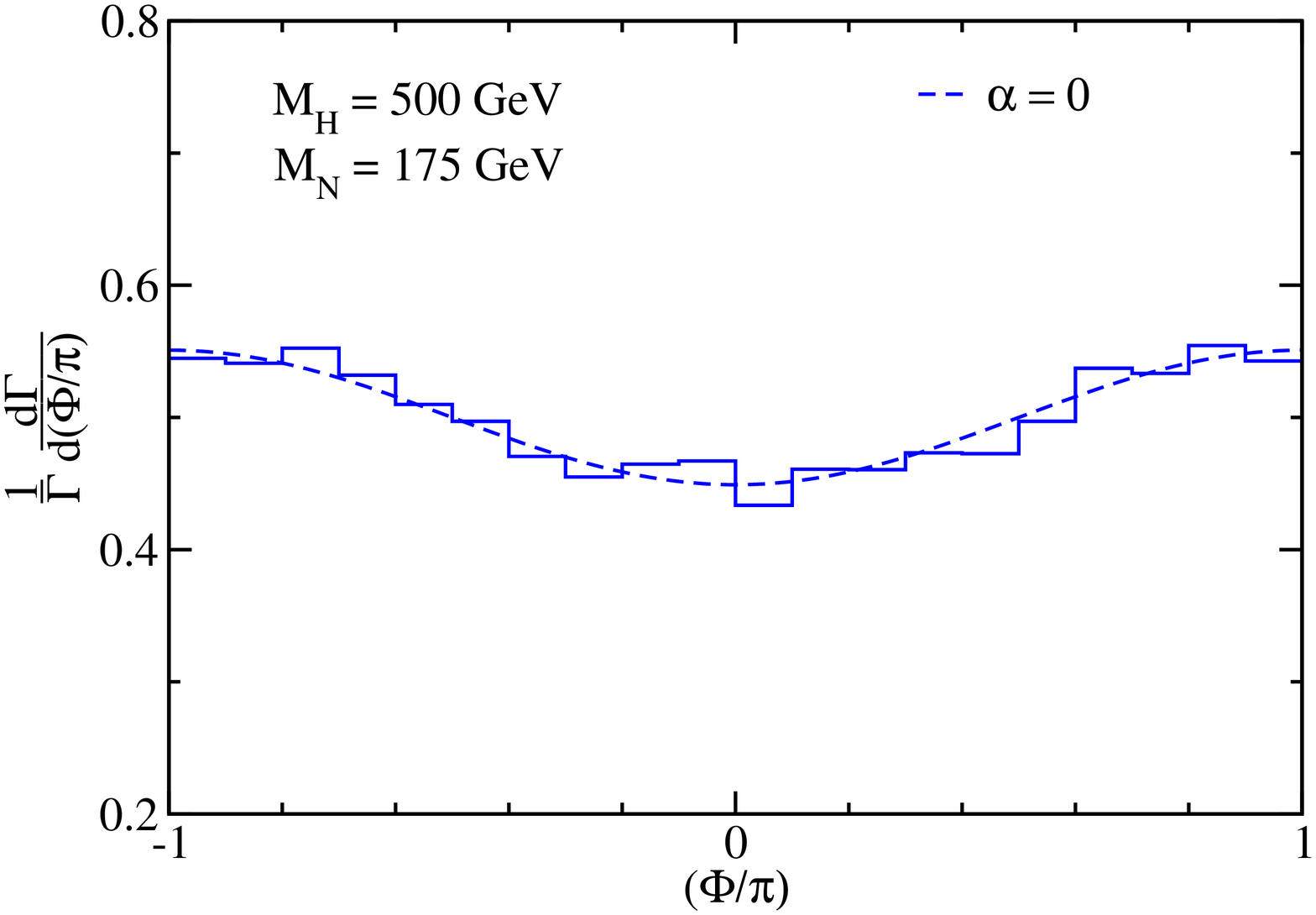}}
\subfigure[]{\includegraphics[clip,width=0.45\textwidth]{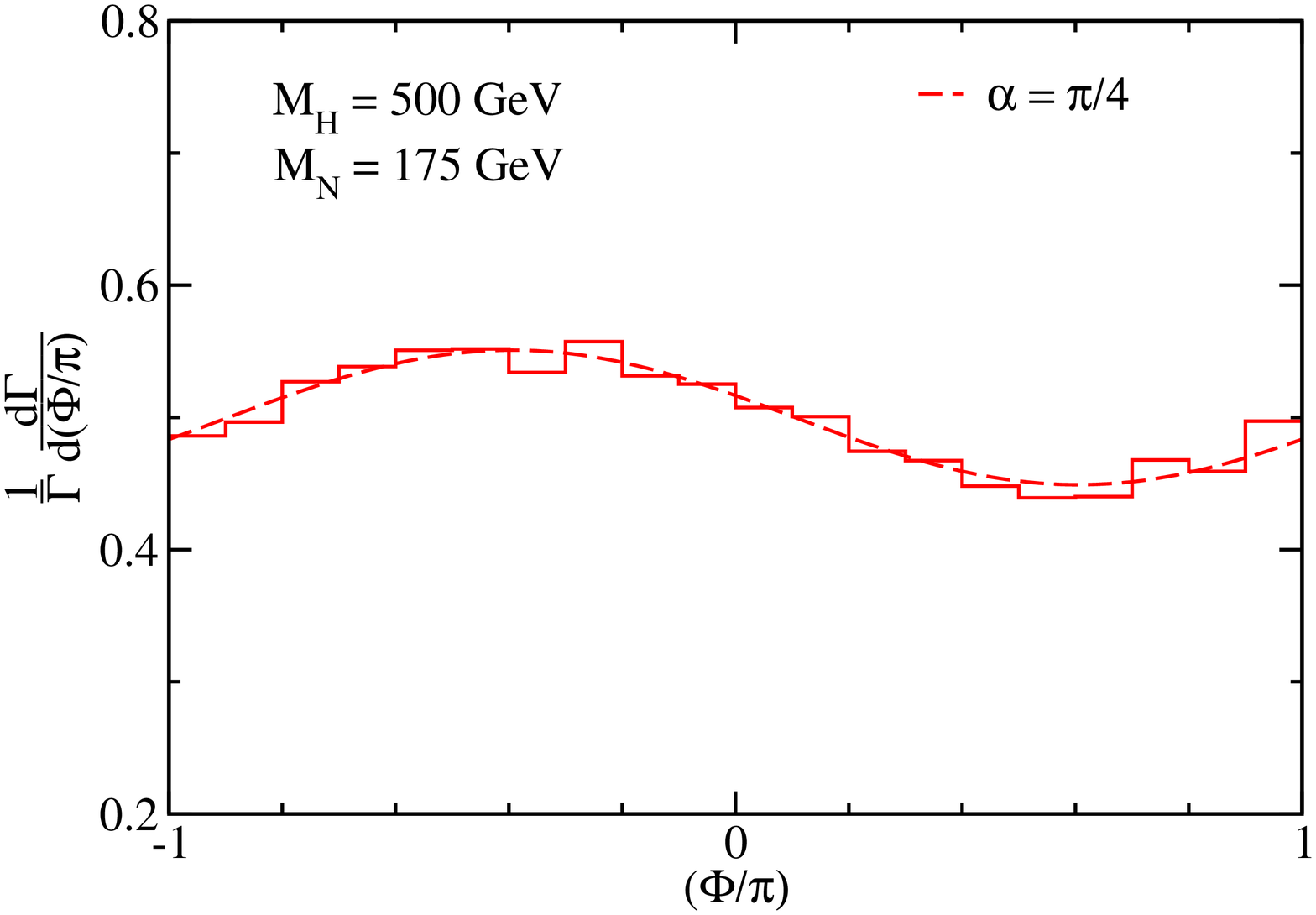}}
\subfigure[]{\includegraphics[clip,width=0.45\textwidth]{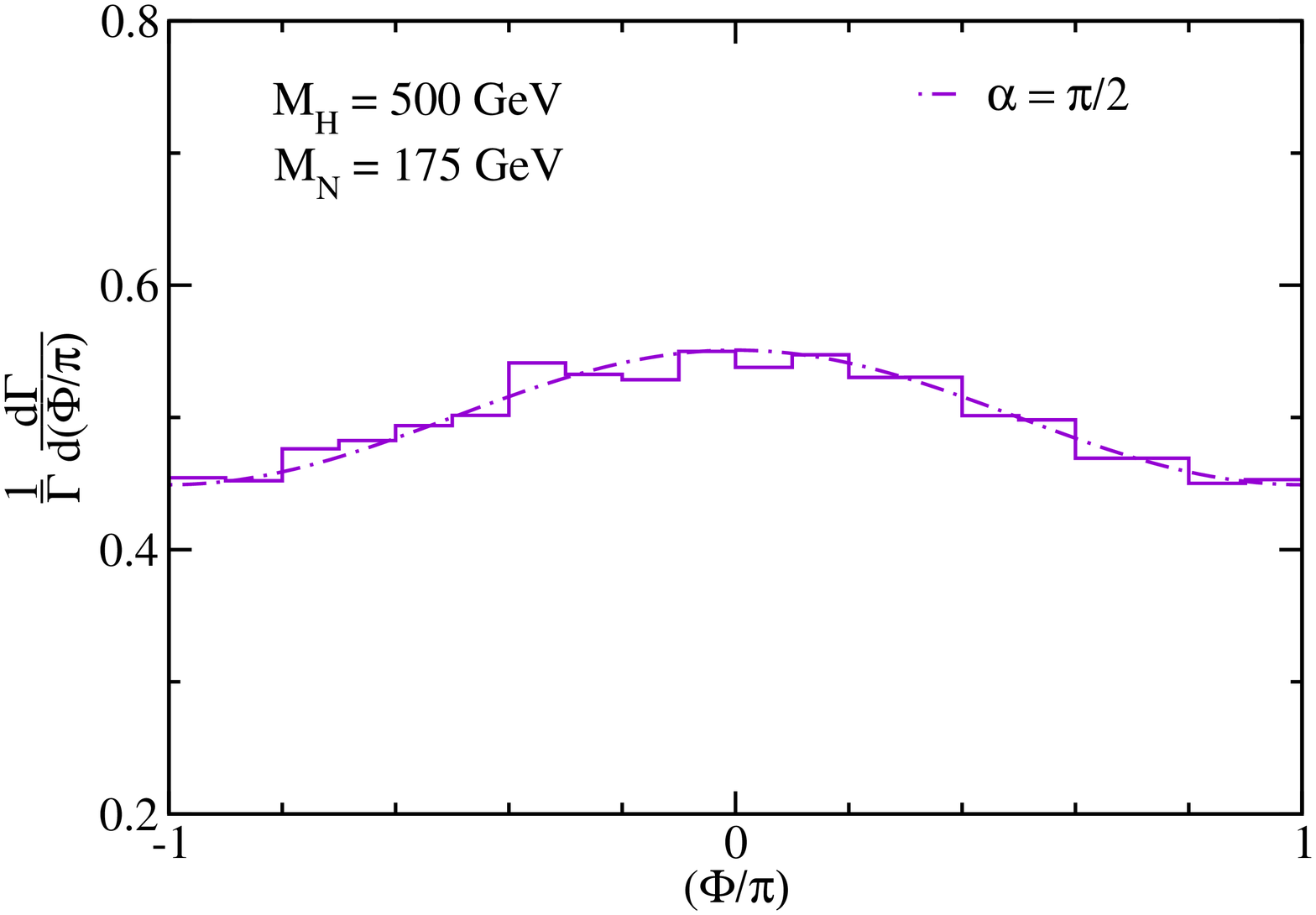}}
\subfigure[]{\includegraphics[clip,width=0.45\textwidth]{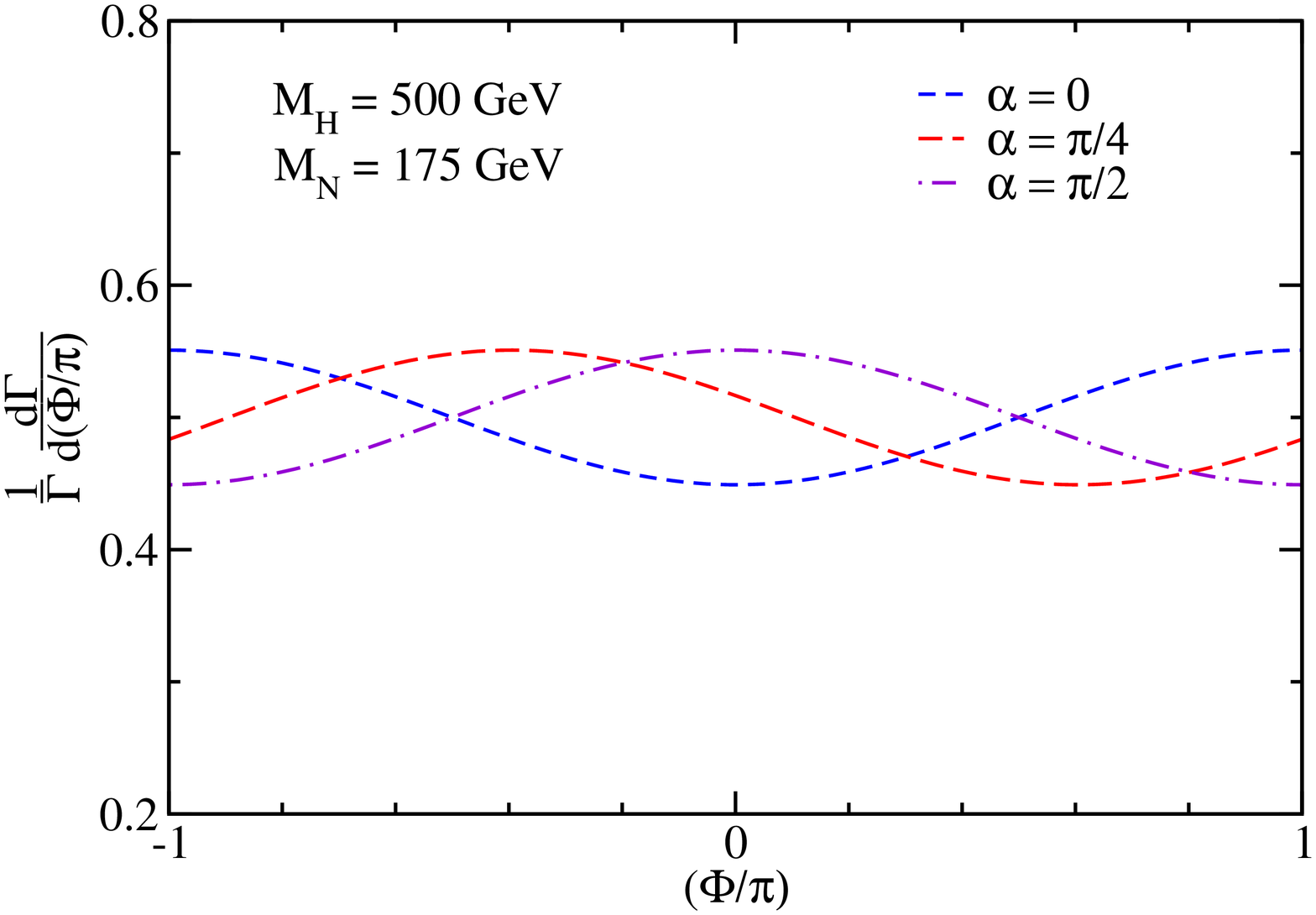}}
\caption{Angular distribution of $\Phi$ for \textbf{opposite-sign} leptons/$W$ bosons using $M_{H}=500$~GeV, $M_{N}=175$~GeV, and a $CP$ phase of (a) $\alpha=0$, (b) $\alpha=\frac{\pi}{4}$, and (c)$\alpha=\frac{\pi}{2}$.  The solid curves give MadGraph results and the dashed curves are our analytic predictions. The theory results for the three choices of phase are plotted together in (d).} \qquad \\
\label{fig:175os}
\end{figure}

\begin{figure}[tb]
\subfigure[]{\includegraphics[clip,width=0.45\textwidth]{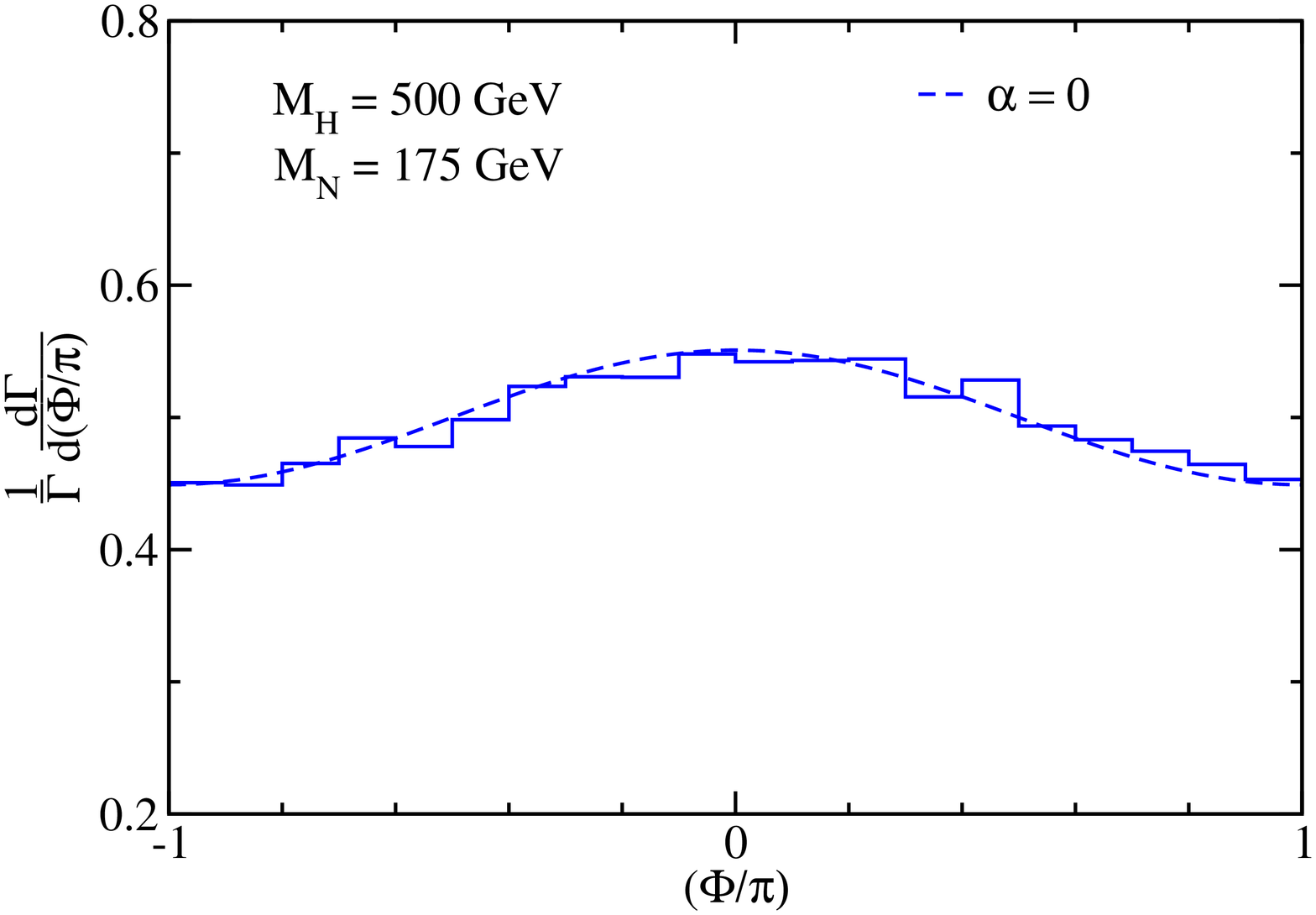}}
\subfigure[]{\includegraphics[clip,width=0.45\textwidth]{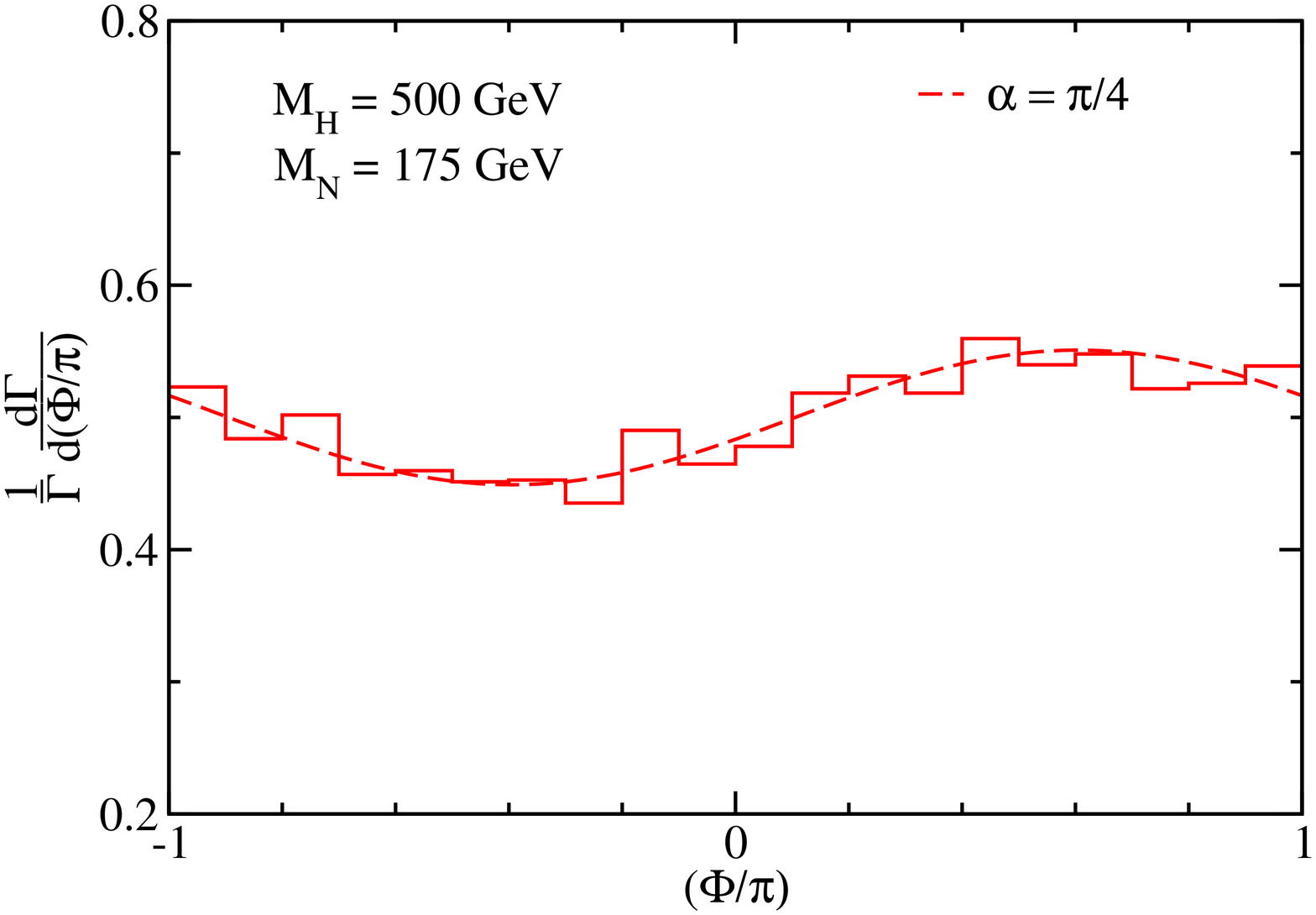}}
\subfigure[]{\includegraphics[clip,width=0.45\textwidth]{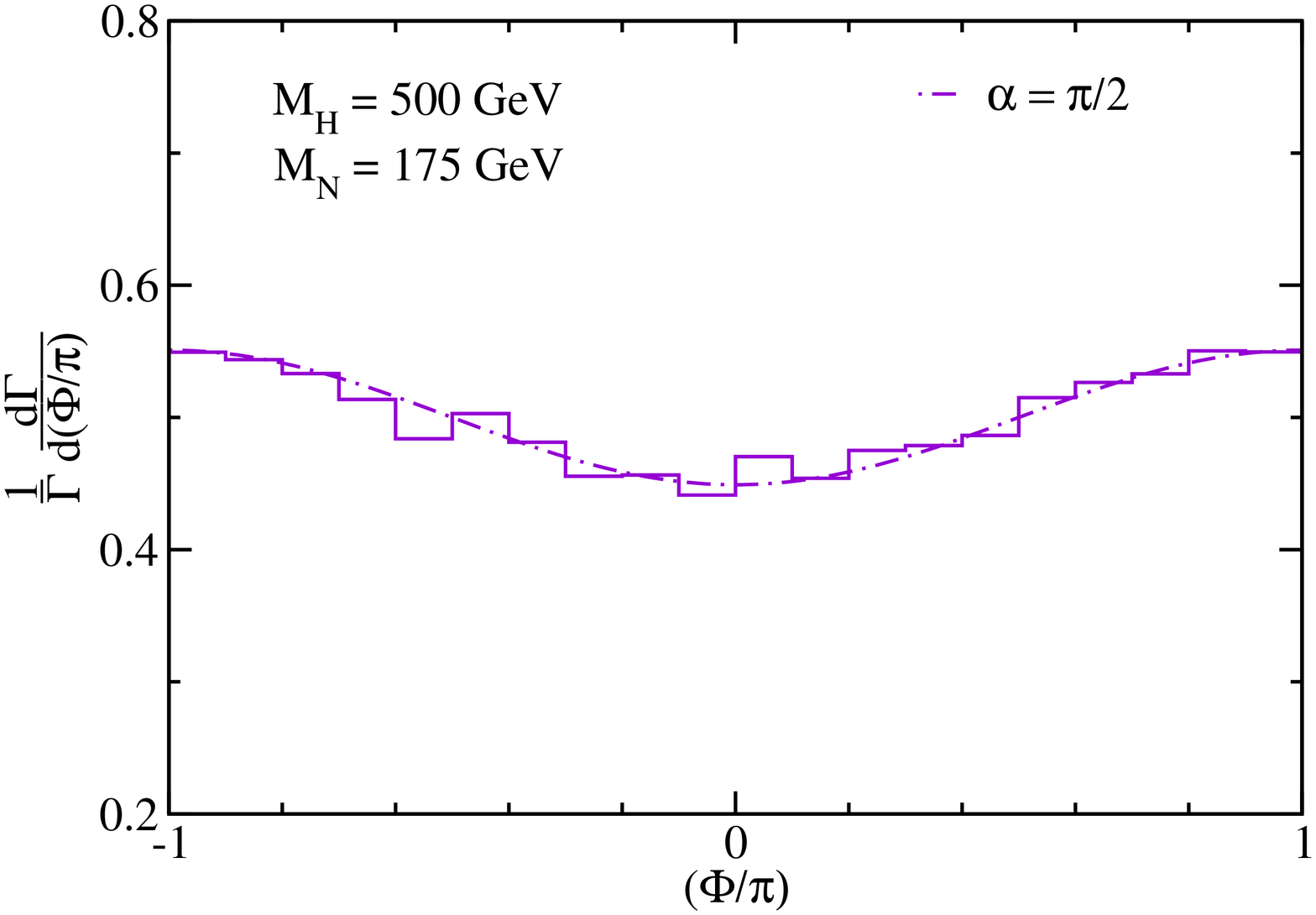}}
\subfigure[]{\includegraphics[clip,width=0.45\textwidth]{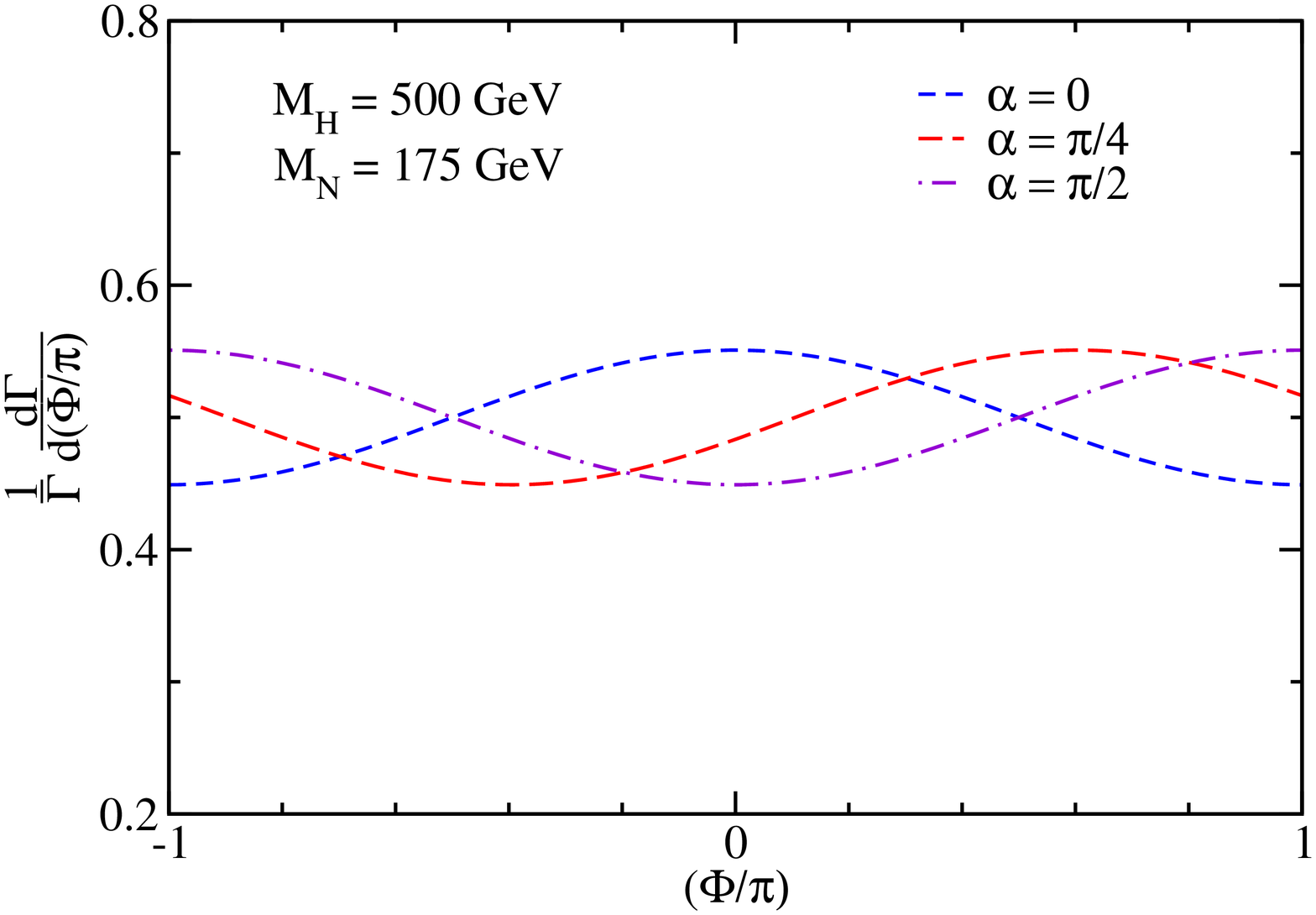}}
\caption{Angular distribution of $\Phi$ for \textbf{same-sign} leptons/$W$ bosons using $M_{H}=500$~GeV, $M_{N}=175$~GeV, and a $CP$ phase of (a) $\alpha=0$, (b) $\alpha=\frac{\pi}{4}$, and (c)$\alpha=\frac{\pi}{2}$.  The solid curves give MadGraph results and the dashed curves are our analytic predictions.  The theory results for the three choices of phase are plotted together in (d).} \qquad \\
\label{fig:175ss}
\end{figure}

We perform a simple check of our results by implementing a model with a heavy Majorana neutrino (with couplings to the Higgs and $W$ boson as described above) in MadGraph/MadEvent v4.4.57 \cite{Alwall:2007st}.  To estimate its decay width, we assume the heavy neutrino decays entirely through $W$ bosons with $V_{lN} \sim 10^{-3}$, such that $\Gamma_{N} \approx 2 |V_{lN}|^{2} \Gamma_{t} \sim 10^{-6}$~GeV (where $\Gamma_{t}\approx 1.5$~GeV is the width of the top quark).  We look at the decays $H \to N N \to l^{+} l^{-} W^{+} W^{-}$ and  $H \to N N \to l^{\pm} l^{\pm} W^{\mp} W^{\mp}$ and in each case generate $50 000$ events in the rest frame of the Higgs boson.  In Figs.~\ref{fig:175os}~(a)-(d) we present the partial decay width as a function of $\Phi$ for the opposite-sign signal using $M_{H}=500$~GeV and $M_{N}=175$~GeV.  This is plotted for a $CP$-even Higgs boson ($\alpha=0$), a $CP$-odd Higgs boson ($\alpha = \pi / 2$), and mixed-$CP$ state ($\alpha = \pi / 4$).  In all cases, the MadGraph results show the expected behavior.  The same results are given for the same-sign signal in Figs.~\ref{fig:175ss}~(a)-(d).  The analytic curves match the MadGraph results quite well and demonstrate the overall phase difference of $\pi$ with the opposite-sign results, as described above.

In both cases, it is important to properly pair the leptons and $W$ bosons to reconstruct two on-shell neutrinos; in the opposite-sign case, the charge of the $W$ bosons may be ambiguous if they both decay hadronically and in the same-sign case there is a fundamental ambiguity even if the charges are known.  We followed a very simple procedure here: as we know each event is derived from the decay of two equal mass intermediate states with a very narrow width, we take the combination that minimizes $|(l_{1}+W_{i})^{2} - (l_{2}+W_{j})^{2}|$.  This works rather well for our MadGraph-level results; however, due to smearing effects and the inherent difficulties that will arise in reconstructing the Higgs rest frame, more elaborate techniques will likely be necessary in practice.

\begin{figure}[tb]
\subfigure[]{\includegraphics[clip,width=0.45\textwidth]{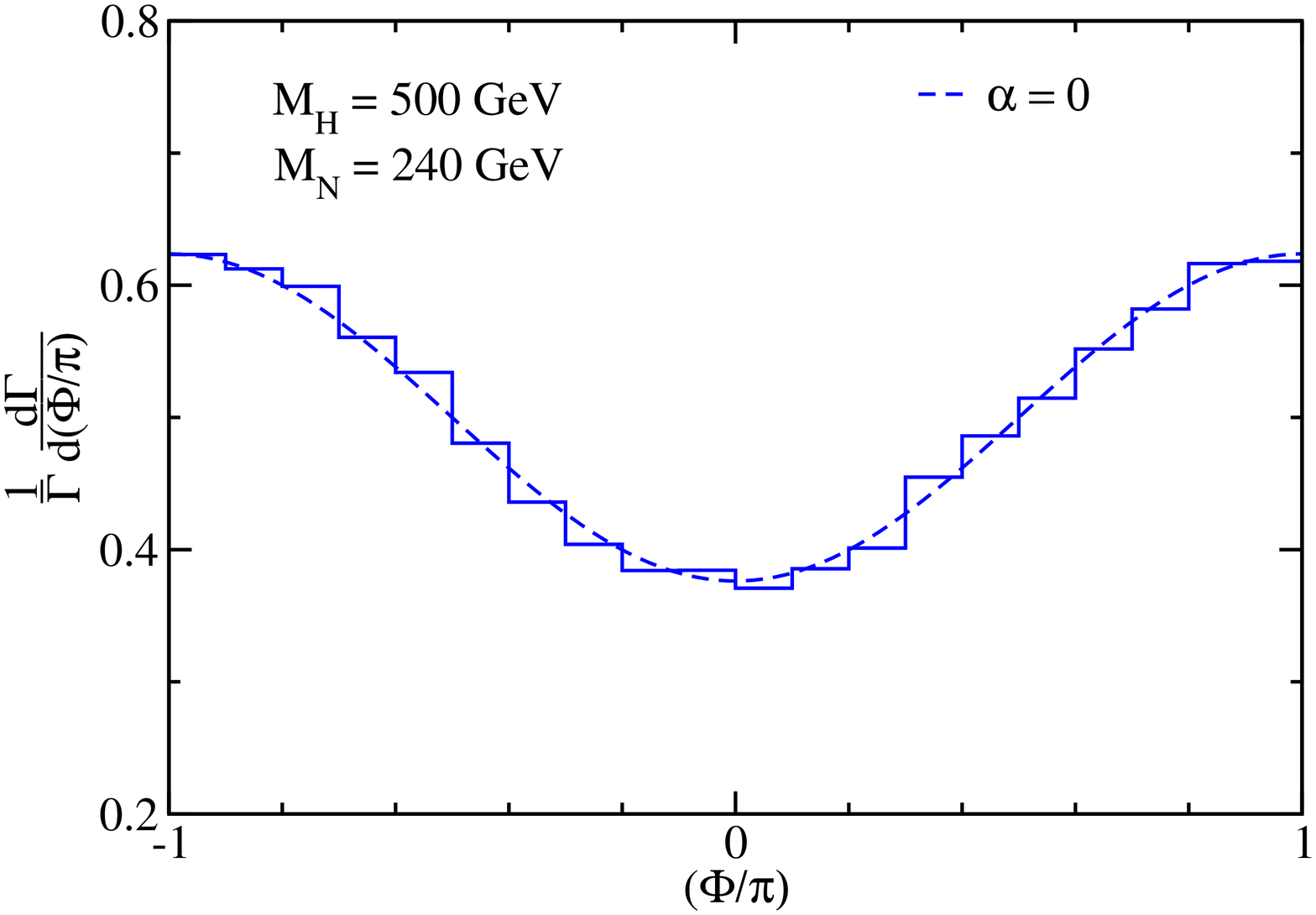}}
\subfigure[]{\includegraphics[clip,width=0.45\textwidth]{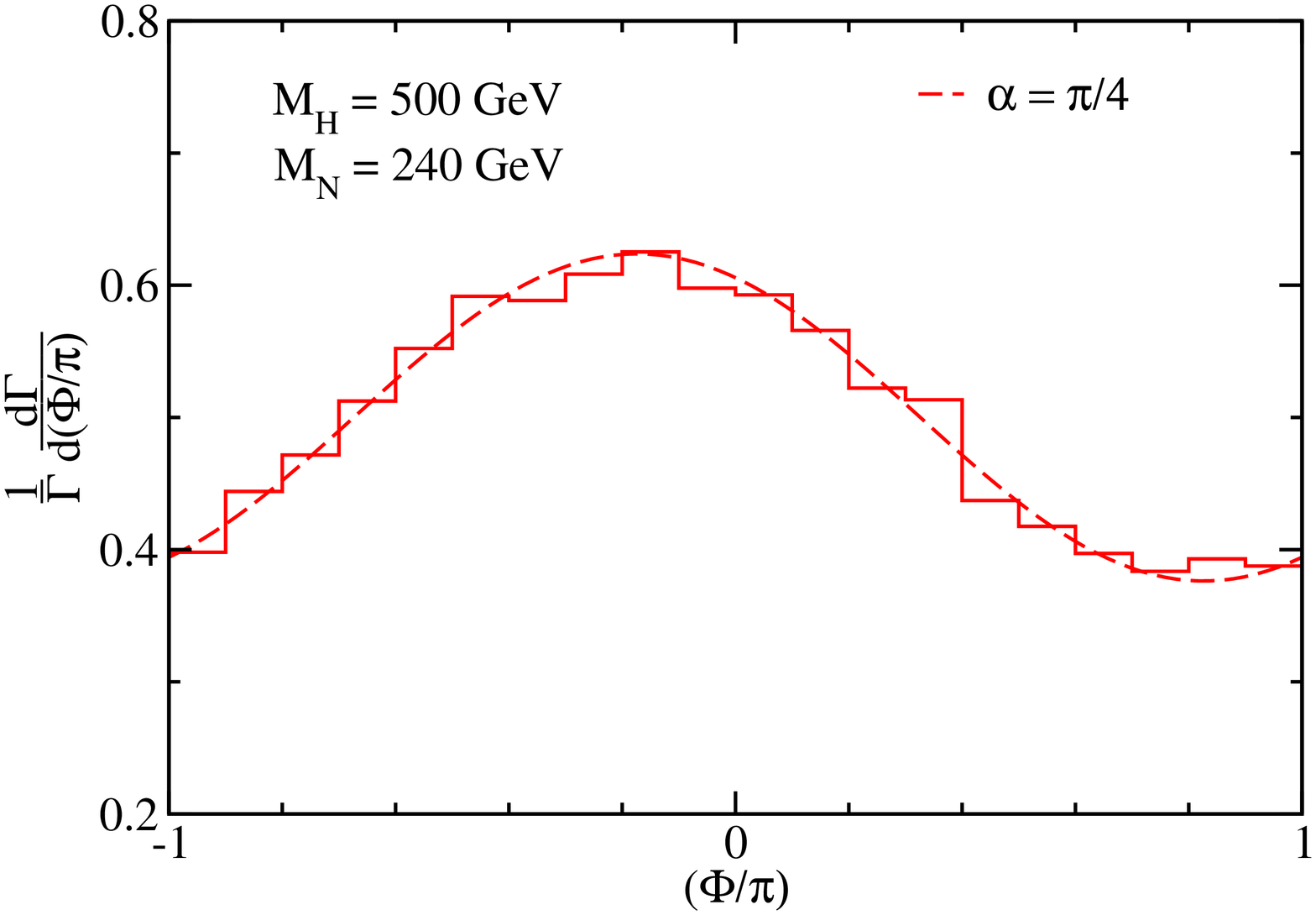}}
\subfigure[]{\includegraphics[clip,width=0.45\textwidth]{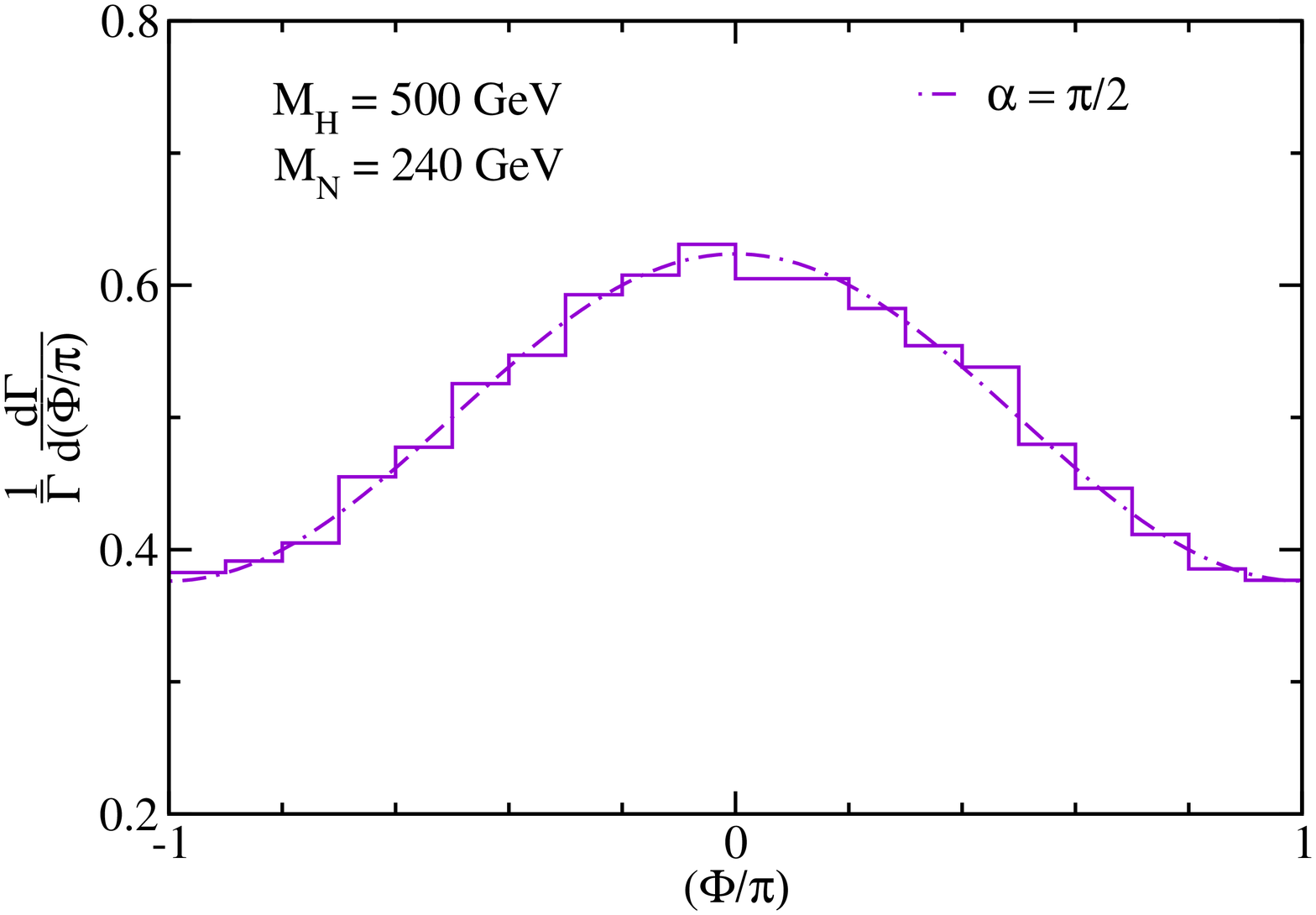}}
\subfigure[]{\includegraphics[clip,width=0.45\textwidth]{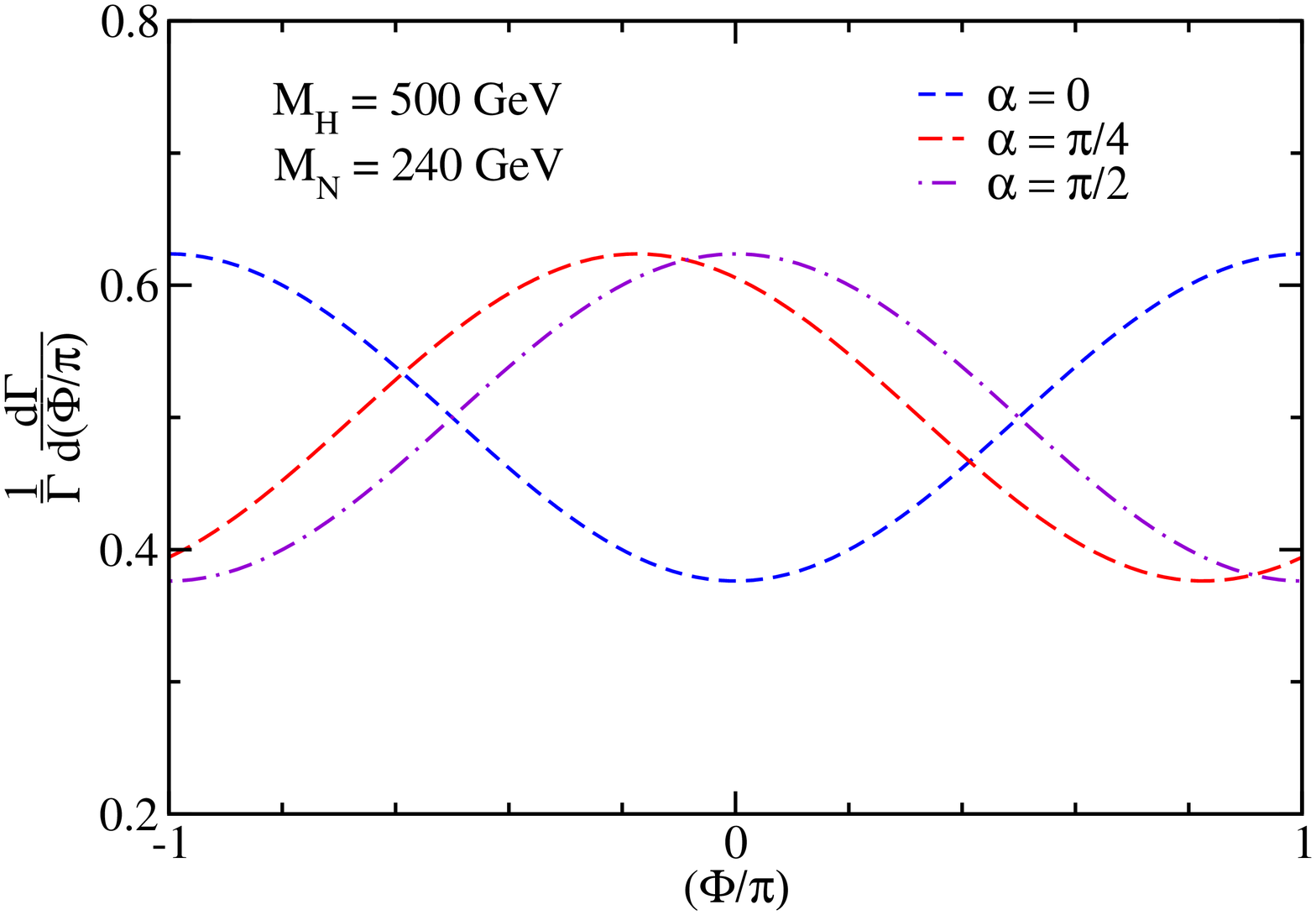}}
\caption{Angular distribution of $\Phi$ for \textbf{opposite-sign} leptons/$W$ bosons using $M_{H}=500$~GeV, $M_{N}=240$~GeV and a $CP$ phase of (a) $\alpha=0$, (b) $\alpha=\frac{\pi}{4}$, and (c)$\alpha=\frac{\pi}{2}$.  The solid curves give MadGraph results and the dashed curves are our analytic predictions. The theory results for the three choices of phase are plotted together in (d).} \qquad \\
\label{fig:240os}
\end{figure}

In Figs.~\ref{fig:240os}~(a)-(d) we show, for comparison, results derived from the opposite-sign signal for a heavy neutrino with mass just below production threshold, $M_{N} = 240$~GeV.  This demonstrates both the enhancement to the relative amplitude due to the small value of $M_{W} / M_{N}$ as well as a phase-shift $2 \chi \approx \pi $ in the case of $\alpha \ne  0$ with near-threshold production.

%%%%%%%%%%%%%%%%%%%%%%%%%%%%
\section{Conclusions} \label{sec:conclusions}
%%%%%%%%%%%%%%%%%%%%%%%%%%%%

We have demonstrated that the $CP$ phase of a Higgs boson state in an extended Higgs sector may, in principle, be determined at the LHC by looking at angular correlations in the signals $H \to N N \to l^{+} l^{-} W^{+} W^{-}$ and $H \to N N \to l^{\pm} l^{\pm} W^{\mp} W^{\mp}$.  This is achieved by looking at the differential partial decay width as a function of $\Phi$, the relative azimuthal angle of the leptons / $W$ bosons about the axis defined by the neutrino momenta in the Higgs rest frame.  The $CP$ phase of the Higgs boson, $\alpha$, introduces a phase-shift $2 \chi = 2 \tan^{-1} ( \tan(\alpha) / \beta_{N})$ in the $\cos\Phi$ dependence of the decay width, where $\beta_{N}$ is the velocity of the neutrinos in the Higgs rest frame.  Measurements of this phase shift will allow a direct determination of the $CP$ nature of the Higgs boson with this signal.  

We have also verified that in the context of a fourth generation with heavy quarks, the process $p p \to H \to N N$ can have a cross section of $\approx 100$~fb - $5000$ fb at the LHC if the Higgs boson is heavy enough for on-shell decays to heavy neutrinos with mass $M_{N} > 100$~GeV.  This process, including the subsequent $N$ decays, may therefore be observable above background, particularly in the case of the same-sign signal, which may allow for the azimuthal distributions to be measured at the LHC. 

%%%%%%%%%%%%%%%%%%%%%%%%%%%%
\section{Acknowledgments}
%%%%%%%%%%%%%%%%%%%%%%%%%%%%

The authors would like to thank I. Lewis for many helpful discussions.  W.-Y.K. and B.Y. thank BNL for hospitality.  This work was supported in part by the U.S. Department of Energy under Grants No.~DE-FG02-95ER40896 and No. DE-FG02-84ER40173 and in part by the Wisconsin Alumni Research Foundation.

\newpage
%%%%%%%%%%%%%%%
\bibliographystyle{h-physrev}
%\newpage
\bibliography{ms.bbl}
%%%%%%%%%%%%%%%s
\newpage

\end{document}